\documentclass[useAMS,usenatbib,usegraphicx]{mn2e}
\usepackage{epsfig}
\usepackage{amsmath} 
\usepackage{rotating}           
\usepackage{color}                

\usepackage{times}

\def\kms{km ${\rm s}^{-1}$}

\def\ch2{$\chi^2$}

\def\Mo{M$_\odot$}

\def\kms {\hbox{${\rm km\ s}^{-1}$}}

\def\ccm {$\hbox{{\rm cm}}^{-3}$}    
\def\scm  {$\hbox{{\rm cm}}^{-2}$}    
\def\arcsec {\hbox{$^{\prime\prime}$}}

\def\arcmin {\hbox{$^{\prime}$}}

\def\MOLH {\hbox{${\rm H}_2$}}  

\def \AL {$\alpha $}     
\def \HI {H{\sc \,i}}
\def \WpHz {W Hz$^{-1}$}

\def\lapp{\ifmmode\stackrel{<}{_{\sim}}\else$\stackrel{<}{_{\sim}}$\fi}
\def\gapp{\ifmmode\stackrel{>}{_{\sim}}\else$\stackrel{>}{_{\sim}}$\fi}

\title[Hydrogen in high redshift radio sources]{A survey for the missing hydrogen in high redshift radio sources} 
\author[S. J. Curran et al.]{S. J. Curran$^{1,2}$\thanks{E-mail:
sjc@physics.usyd.edu.au}, M. T. Whiting$^{3}$,  E. M. Sadler$^{1,2}$  and C. Bignell$^{4}$\\
$^{1}$Sydney Institute for Astronomy, School of Physics, The University of Sydney, NSW 2006, Australia\\
$^{2}$ARC Centre of Excellence for All-sky Astrophysics (CAASTRO)\\
$^{3}$CSIRO Astronomy and Space Science, PO Box 76, Epping NSW 1710, Australia\\
$^{4}$National Radio Astronomy Observatory, P.O. Box 2, Rt. 28/92 Green Bank, WV 24944-0002, USA}

\begin{document}

\date{Accepted ---. Received ---; in original form ---}

\pagerange{\pageref{firstpage}--\pageref{lastpage}} \pubyear{2012}

\maketitle

\label{firstpage}

\begin{abstract}
  Unlike at lower redshift, where there is a 40\% detection rate, surveys for 21-cm absorption arising within the hosts
  of $z\gapp1$ radio galaxies and quasars have been remarkably unsuccessful.  \citet{cww+08} suggest that this is due to
  the high redshift selection biasing towards the most optically bright objects (those most luminous in the ultra-violet
  in the rest-frame), where the gas is ionised by the active galactic nucleus. They therefore argue that there must be a
  population of fainter objects in which the hydrogen is not ionised and which exhibit a similar detection rate as at
  lower redshifts. In order to find this ``missing'' gas at high redshift, we have therefore undertaken a survey of
  $z\gapp2$ radio sources, selected by optical faintness. Despite having optical magnitudes which indicate that the
  targets have ultra-violet luminosities below the threshold where all of the gas is ionised, there were no detections
  in any of the eight sources for which useable data were obtained. Upon an analysis of the spectral energy
  distributions, ionising photon rates can only be determined for three of these, all of which 
suggest that the objects are {\em above} the highest luminosity of a current 21-cm detection. The possibility that the other five could
  be located at lower photon rates cannot be ruled out, although zero detections out of five is not statistically
  significant. Another possible cause of the non-detections is that our selection biases the sample towards sources
  which are very steep in the radio band, with a mean spectral index of $\left<\alpha\right> = -1.0$, cf. $-0.3$ for
  both the 21-cm detections and UV luminous non-detections. This adds the further possibility that the sources have very
  extended emission, which would have the effect of reducing the coverage by the putative absorbing gas, thus decreasing
  the sensitivity of the observation.
\end{abstract}


\section{Introduction and sample selection}
\label{intro}

Redshifted \HI\ 21-cm provides a probe of the most abundant element in the Universe, through surveys which are not subject to the
flux limitations of optical surveys. In absorption, 21-cm probes the cool component of the neutral gas, which is
the reservoir of raw material ultimately responsible for all star formation. Furthermore, the strength of the absorption depends only upon
the column density of the absorber and the flux of the background source, making the 21-cm transition potentially detectable up to
redshifts of $z\sim50$, where the ionosphere begins to affect the $\lapp30$ MHz radio waves.


Despite this, redshifted 21-cm absorption is currently rare, with only 78 absorbers known at $z\geq0.1$, 80\% of which
are detected at $z\lapp1$. Much of this is due to the past availability of interference free bands at low frequencies, although
there are additional selection effects at play: 
\begin{itemize}
  \item[-] For absorbers {\em intervening} the sight-lines to more distant radio sources, 
the 60\% detection rate at $z\leq1$, compared to 20\% at $z\geq1$, 
 in systems known to have high hydrogen column densities\footnote{Damped Lyman-$\alpha$ absorption systems (DLAs),
with neutral hydrogen column densities of $N_{\rm HI}\ge2\times10^{20}$ \scm.}, can be attributed to the geometry effects introduced by an
expanding Universe, where the coverage of the background flux is systematically lower at redshifts of $z\gapp1$ \citep{cw06,cur12}. 

\item[-] For absorbers {\em associated} with the radio source itself, the detection rate at 
$z\lapp1$ is 40\%, compared to 17\% at $z\gapp1$ \citep{cw10}.
This is
believed to be due to the higher redshifts biasing towards the most ultra-violet
luminous objects, where the intense flux ionises the gas.
Specifically, associated 21-cm absorption has {\em never} been detected where the $\lambda=1216$ \AA\  continuum 
 luminosity of the active galactic nucleus (AGN) exceeds $L_{\rm 1216}\sim10^{23}$ \WpHz\
(\citealt{cww+08} and recently confirmed by \citealt{gd11}).  At these frequencies ($2.47\times10^{15}$ Hz) and
above, the photons have enough energy to excite the hydrogen beyond the ground state, so that it cannot
absorb in 21-cm. \citet{cw12} have extended this to the ionising ($\lambda \leq912$ \AA) radiation and show
that a similar (i.e. $L_{\rm 912}\sim10^{23}$ \WpHz) cut-off also applies. 
The large fraction of non-detections at high redshift are therefore due to the flux limited nature of the optical surveys
selecting the brightest objects,
where the observed-frame optical light is rest-frame ultra-violet, in which all of the gas is
believed to be ionised \citep{cw12}.
\end{itemize}

\begin{figure*}
\centering \includegraphics[angle=-90,scale=0.80]{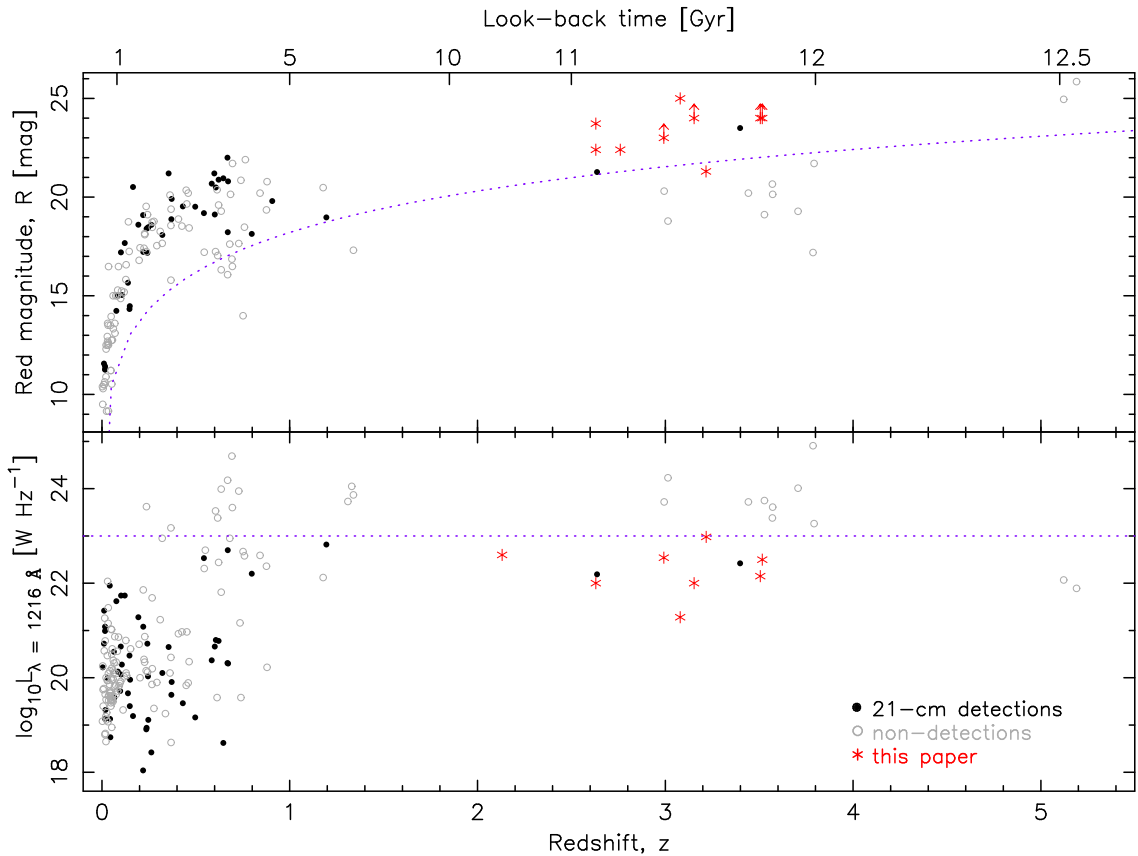}
\caption{The red magnitude (top) and $\lambda = 1216$ \AA\ continuum luminosity (bottom) for the redshifted associated
  21-cm searches (compiled in \citealt{cw10,ace+12}). The filled symbols show the detections and the unfilled symbols
  the non-detections, with the asterisks showing our targets (all of which are non-detections), i.e. the eleven points
  in the top panel and the eight for which useful data were obtained in the bottom panel (with $L_{1216} $ as originally
  estimated and not corrected for extinction, as the remaining points are).  In the top panel the curve shows which $B$ magnitude corresponds to
  $L_{1216}=10^{23}$ \WpHz\ for a spectral slope of $\alpha = -1.5$ (cf. figure 5 of \citealt{cww09}) used to select the
  targets and in the bottom panel the line signifies $L_{1216}=10^{23}$ \WpHz, above which associated 21-cm has never
  been detected. Note that for J0231+3600, at $z=3.079$, $I=25.0$ is shown (Table \ref{obs}).
} 
\label{M-z}
\end{figure*}

In order to find the neutral gas missing from high redshift AGN, we suggest that high redshift 21-cm surveys should be
directed towards the most optically faint sources. We have therefore embarked on an observing campaign of objects
selected by faint optical magnitudes. As discussed in \citet{cwwa11}, our usual source catalogue, the Parkes Half-Jansky
Flat-spectrum Sample (PHFS, \citealt{dwf+97}), yielded only two sources with faint blue magnitudes ($B\gapp22$) in the
90-cm band ($z=3.09 - 3.63$)\footnote{Applying the $B-z$ curve (figure 5 of \citealt{cww09}), gives $B\gapp22$ for
  $L_{\rm 1216}\lapp10^{23}$ \WpHz\ at these redshifts.}, both of which have been previously searched for in 21-cm
\citep{cww+08}. We therefore compiled all available radio catalogues, which give both redshifts and magnitudes, and
selected those with $B, V, R~{\rm or}~I$ magnitudes which indicate that $L_{1216}\lapp10^{23}$ \WpHz\ at the given
redshift (Fig. \ref{M-z}).
We further shortlisted
those in which 21-cm is redshifted into the 290--395 MHz band of the Green Bank Telescope (GBT) or the UHF-low band
(250--460 MHz) of the Westerbork Synthesis Radio Telescope (WSRT) and which reached sufficiently high elevations at
these locations, while having flux densities estimated to be $\gapp0.2$ Jy at the redshifted 21-cm
frequency.\footnote{Except for J0617+5012 for which no fluxes were available from the NASA/IPAC Extragalactic Database,
  although we report 1.28 Jy at 342 MHz here.} This gave eleven sources which we observed (Table~\ref{obs}), eight
of which were not completely ruined by RFI, and as seen from Fig.~\ref{M-z}, all are believed to be below the critical luminosity.
For the $\approx40\%$ detection rate at $L_{1216}\lapp10^{23}$ \WpHz, we therefore expect approximately three detections,
although there were none. We discuss possible reasons for the exclusive non-detections in this paper.

\section{Observations and data reduction}
\label{observations}


\subsection{Green Bank observations}
\label{gbo}

Each of the sources targetted with the Green Bank Telescope were observed for a
total of two hours on source,  with the aim of reaching an r.m.s. noise level of $\lapp5$ mJy per 10 \kms\ channel (for $T_{\rm sys}
= 70$ K). This gives a $3\sigma$ optical depth limit of
$\tau\approx0.03$ per channel for a flux density of 0.5 Jy, i.e. a sensitivity to $N_{\rm HI}\approx3\times10^{17}(T_{\rm spin}/f)$ \scm,
the lower limit for most of the published searches (see figure 4 of \citealt{cww+08}). We used the {\tt Rcvr\_342} receiver backed with the GBT
Spectrometer in narrow bandwidth, high resolution mode in 
$4\times12.5$ MHz bands, 
while maintaining a redshift
coverage of $\Delta z \approx \pm0.06$ in each band. Each source was observed in two orthogonal linear polarisations
which, after flagging of bad data, were averaged together (using the {\sc gbtidl} software):\\
{\bf NVSS\,J012142+132058} was observed at 314.53 MHz on 17 June 2010, over 4096 channels, giving a spacing of 3.052 kHz (2.92 \kms). After 
flagging of the worst RFI affected scans, 0.6
hours of data remained with an average system temperature of $T_{\rm sys} = 90$ K, although some spikes were still present in the bandpass  (Fig. \ref{spectra}).\\
{\bf  NVSS\,J020510+224250} was observed at 315.23 MHz on 31 August 2010, over 4096 channels, giving a spacing of 3.052 kHz (2.91 \kms).
$T_{\rm sys} = 87$ K  and the {\sf YY} polarisation exhibited negative fluxes for the whole observations and after flagging out the worst RFI affected on the {\sf XX} polarisation
0.68 hours of data remained.\\
{\bf  NVSS\,J023111+360027} was observed at 348.22 MHz on 1 September 2010, over 8196 channels, giving a spacing of 1.526 kHz (1.31 \kms).
After flagging, 0.74 hours of data remained, although all of the scans exhibited negative fluxes. $T_{\rm sys} = 75$ K   and between the RFI spikes, the r.m.s. noise levels were low
(4 mJy per 10 \kms), although the spikes raised the mean to 189 mJy over the observed range.\\
{\bf WN\,J0617+5012} was observed at 342.02 MHz on 6 July 2010, over 8196 channels, giving a spacing of 1.526 kHz (1.34 \kms). The mean system temperature
was 76 K and after flagging, 0.63 hours of data remained. There is an ``absorption feature'' close to the expected redshift of $z= 3.153$ (fitted by
two narrow Gaussians at an offset of 220 and 241 \kms\ from 0 \kms,\ defined by 342.02 MHz), which is present on all four sub-averages of the data.
However, this is also true for the two ``emission features'' at $\approx -1500$ and $-1100$ \kms\ and so we must attribute these to RFI. Again,
the regions between the spikes had low  r.m.s. noise levels (5 mJy per 10 \kms), although in Table \ref{obs} we quote the value over 337.56 --343.17 MHz,
were the spikes are fairly weak.\\
{\bf PKS\,0742+10} was observed at 390.30 MHz on 13 August 2010, over 16\,384 channels, giving a spacing of 763~kHz (0.59 \kms).
After flagging of the worst RFI affected scans, 
0.6 hours of data remained with an average system temperature of $T_{\rm sys}
= 81$ K. However, intermittent spikes remained, meaning that the r.m.s. noise level quoted in Table \ref{obs}
only applies to the clean regions between these, with a spike being present over  $z = 2.6302-2.6309$
(391.20--391.27 MHz), very close to the emission redshift of the source.\\
{\bf FIRST\,J074729.3+365438} was observed at 355.81 MHz on 7 June 2010, over 16\,384 channels, giving a spacing of
763~kHz (0.66 \kms).  The mean system temperature was 63 K and 0.8 hours of data remained after flagging the
integrations affected most by RFI. As seen in Fig. \ref{spectra},
the bandpass was still dominated by spikes and so in Table \ref{obs} we quote the r.m.s. noise level over the relatively clear 355.78--357.15 MHz.\\
{\bf NVSS\,J092022--071217} was observed at 377.77 MHz on 9 September and 29 December 2010, over 16\,384 channels,
giving a spacing of 763~kHz (0.61 \kms).  The mean system temperature was 65 K and flagging of badly affected data gave
a total integration time of 0.63 hours. Even so, the bandpass was still dominated by spikes with RFI close to the
expected frequency of the redshifted 21-cm line preventing us from assigning a
limit.\\
{\bf B2\,1121+31B} was observed at 336.83 MHz in three sessions spanning from 6 July to 5 August 2010. The band was split
over 4096 channels, giving a spacing of 3.052 kHz (2.73 \kms).  The mean system temperature was 62 K and RFI was minimal, with
0.90 hours of
data remaining after flagging the worst of this.  Despite the absence of severe RFI and the flat bandpasses, like J0205+2242 each scan
exhibited a negative flux, in both polarisations, over the whole run.  Negative fluxes can often caused by a region of high
confusion leading to the off-measurement
being stronger than the
source. At 337 MHz, the half-power beam width of the GBT is $\sim40\arcmin$, within which the NASA/IPAC Extragalactic Database
lists 17\,000 sources, 140 of which are classified as radio sources. Therefore confusion within the beam/sidelobes/off-position
leading to a negative flux is a possibility. Although the flux levels may be 
not be properly calibrated, the data themselves are believed to the reliable \citep{rie12}, as is apparent from the consistency
between scans for both J0205+2242 and 1121+31B.

\subsection{Westerbork observations}

\begin{figure*} 
\vspace{12.0cm}  
\includegraphics{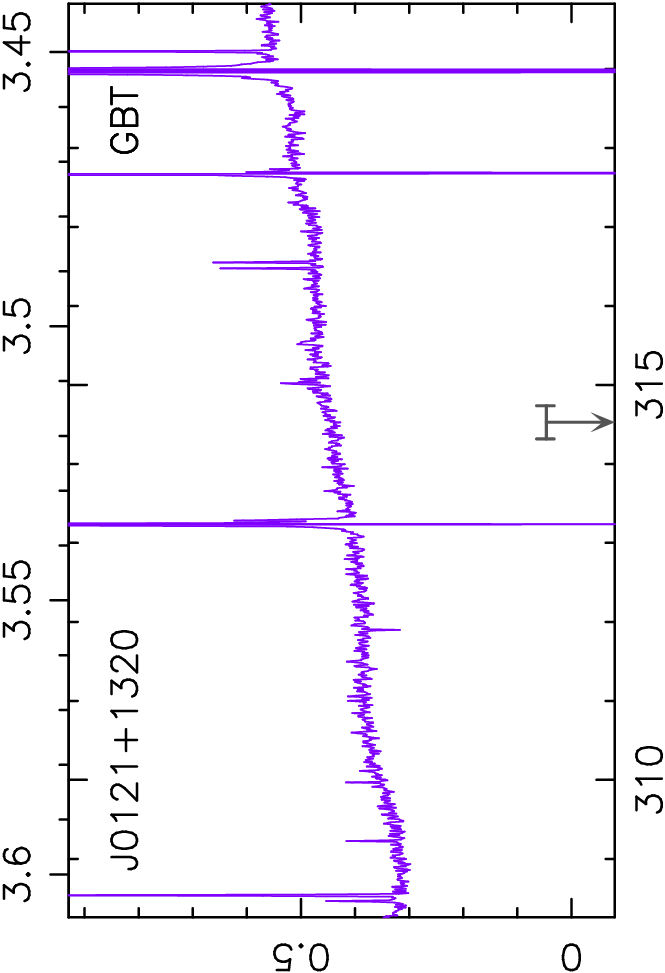}
\includegraphics{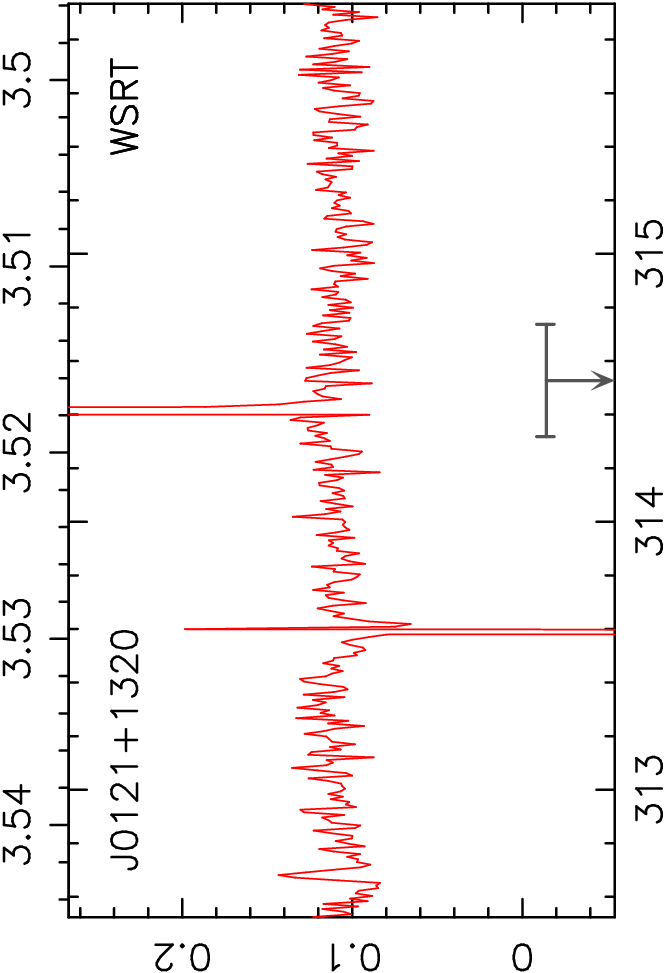}
\includegraphics{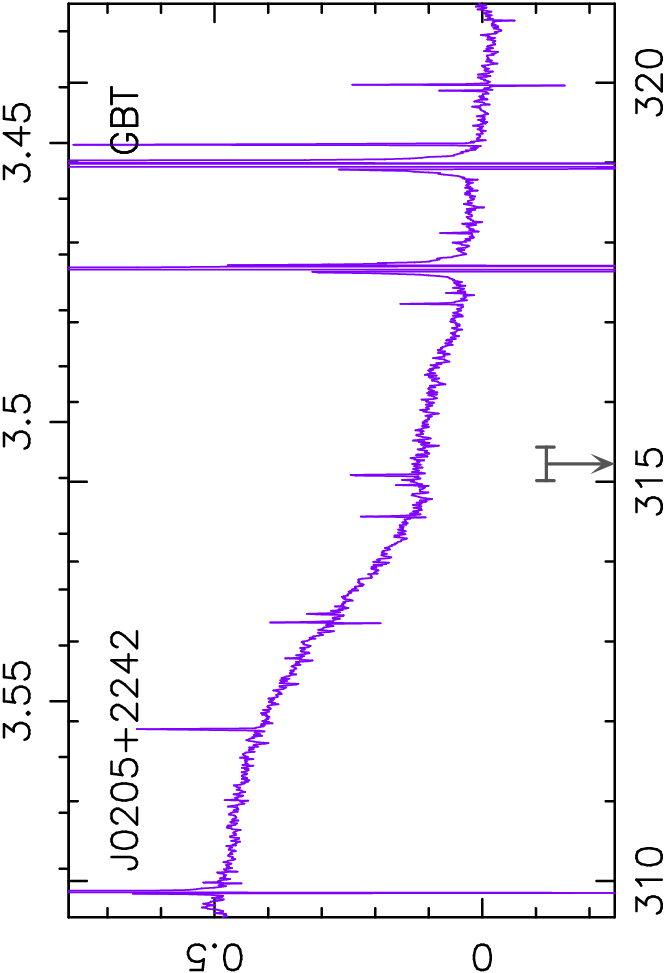}
\includegraphics{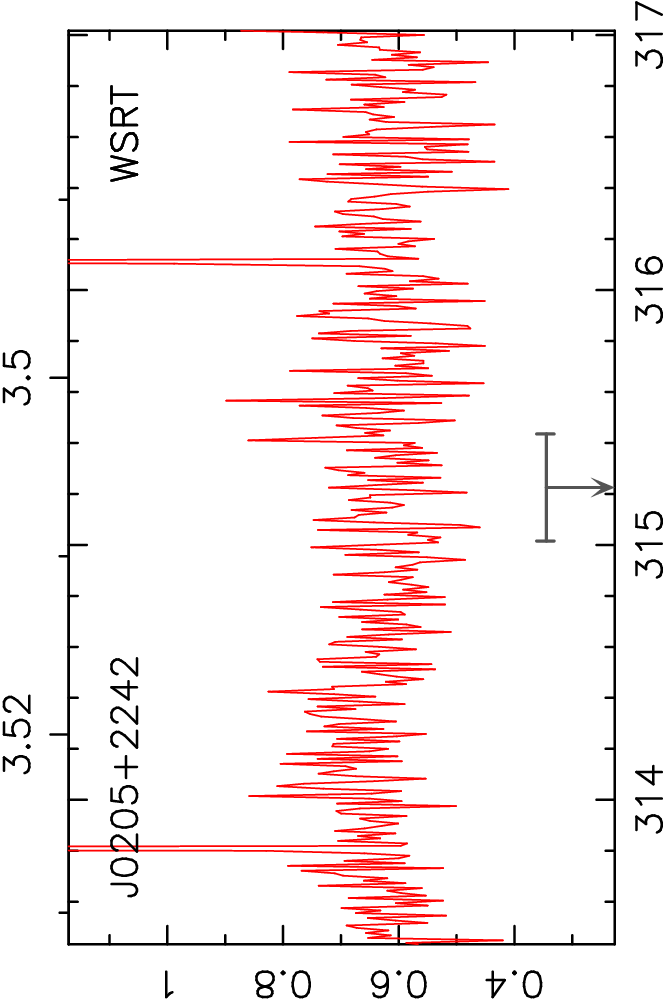}
\includegraphics{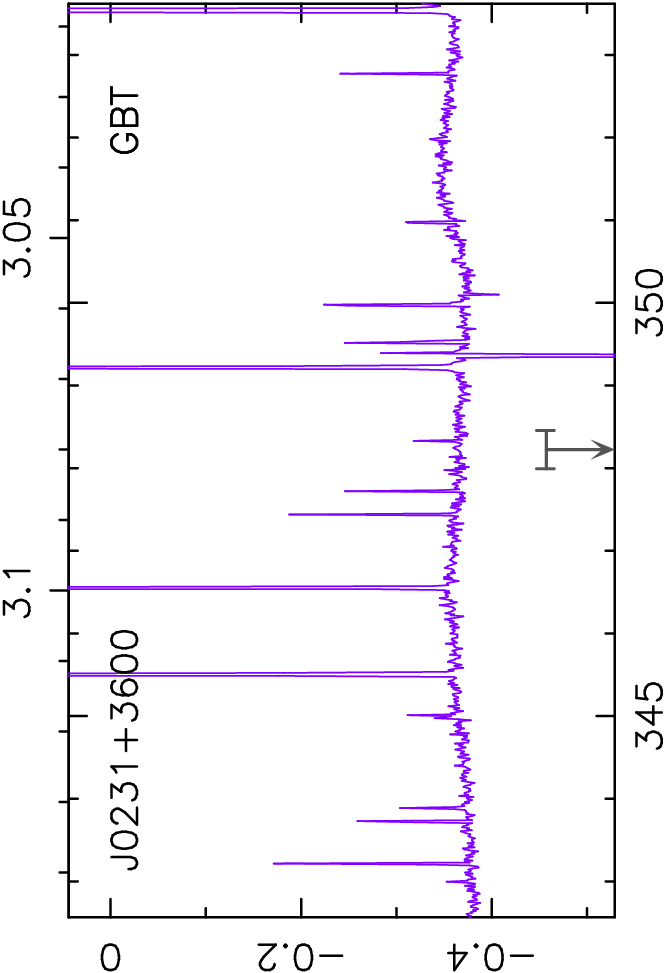}
\includegraphics{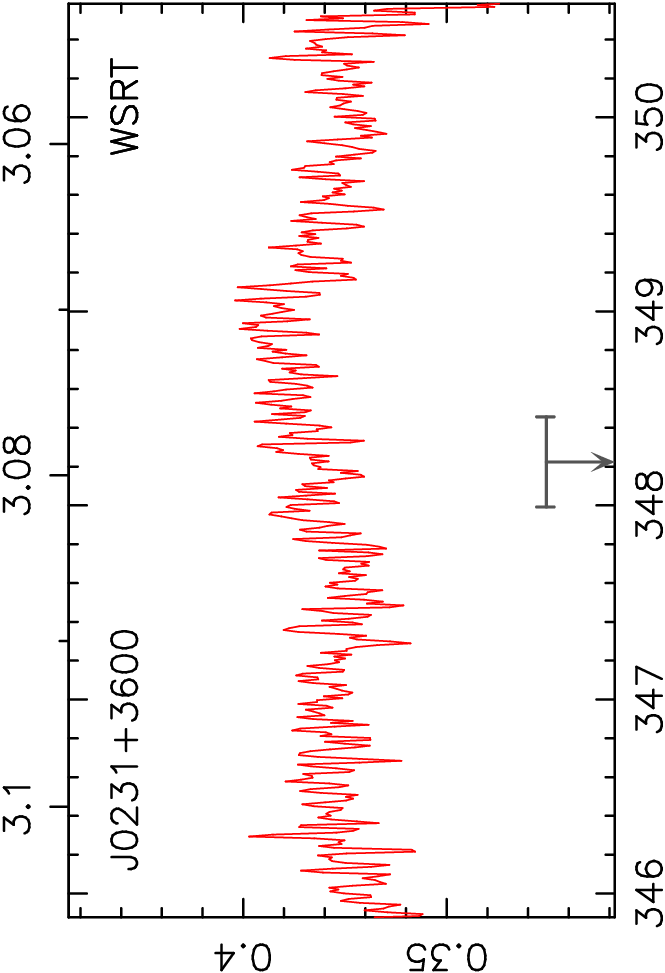}
\includegraphics{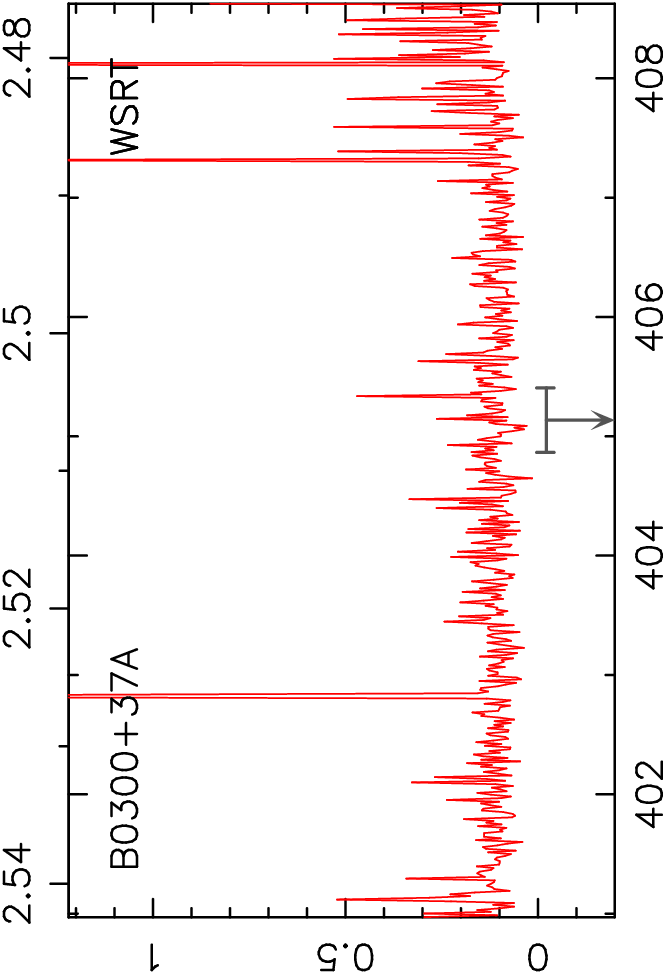}
\includegraphics{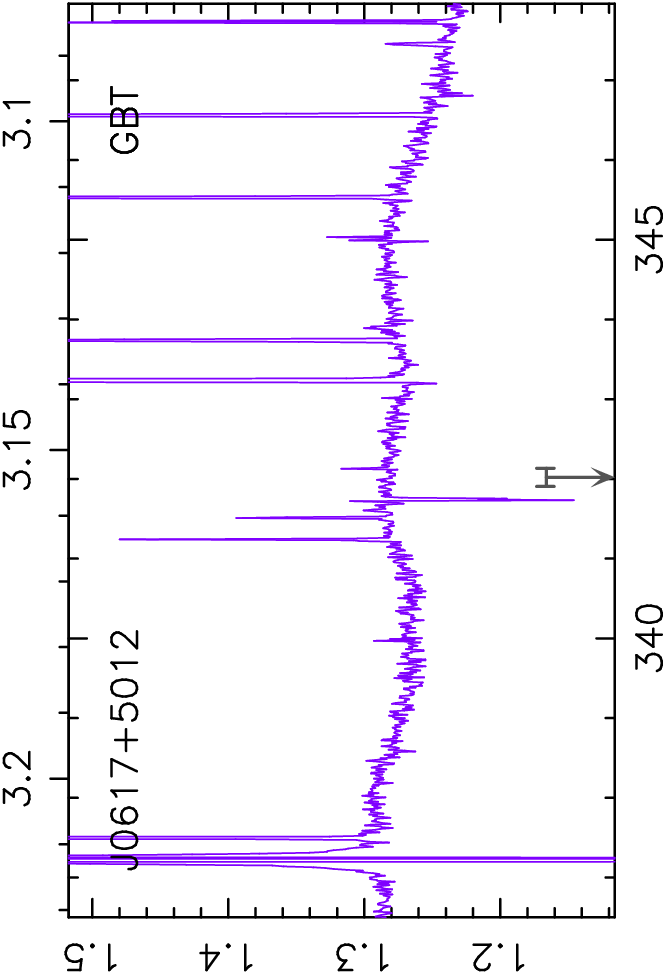}
\includegraphics{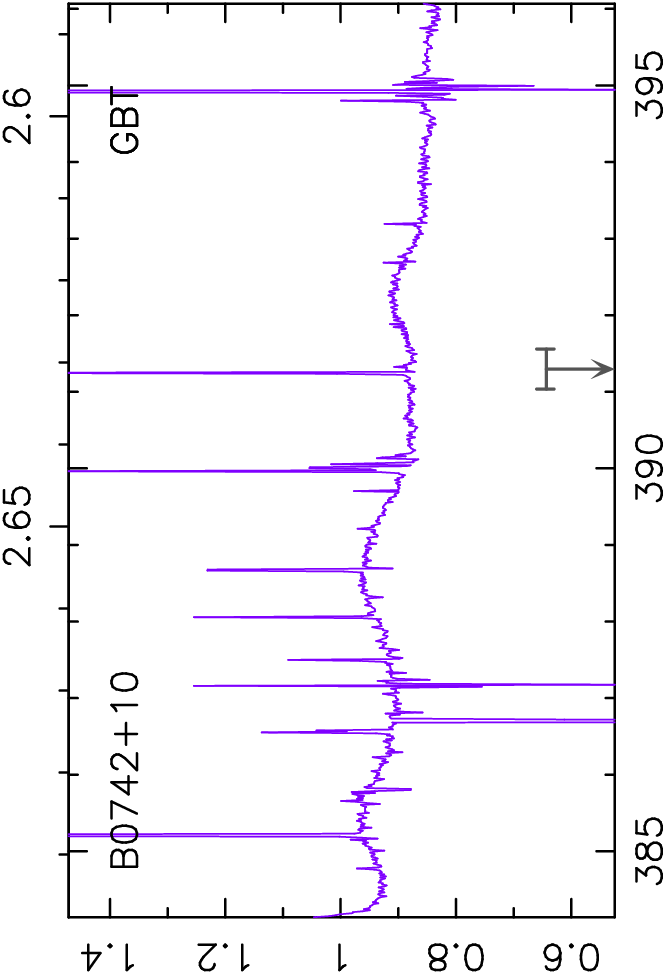}
\includegraphics{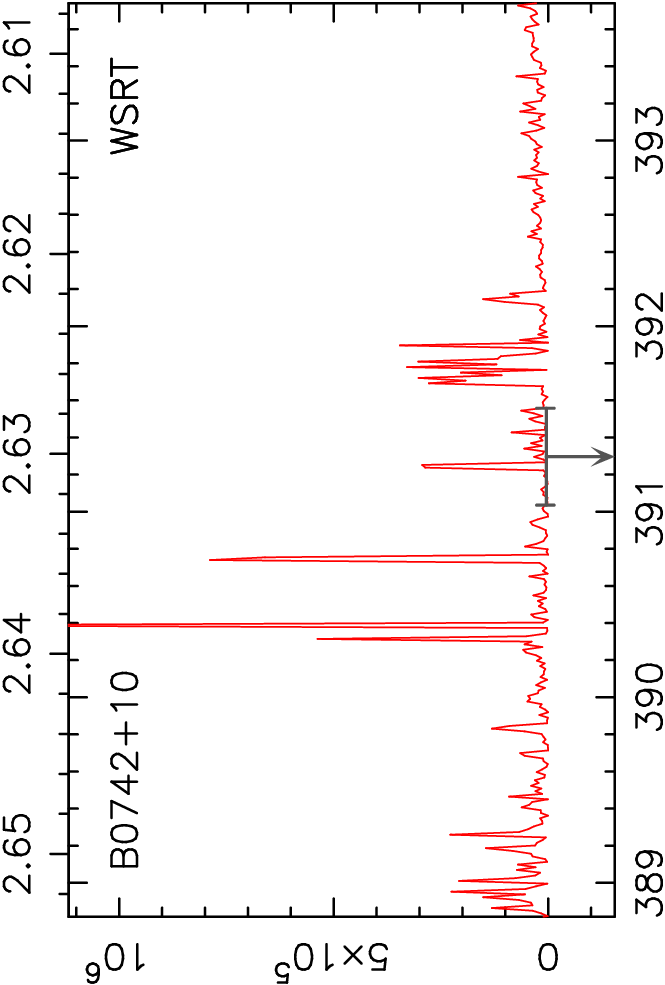}
\includegraphics{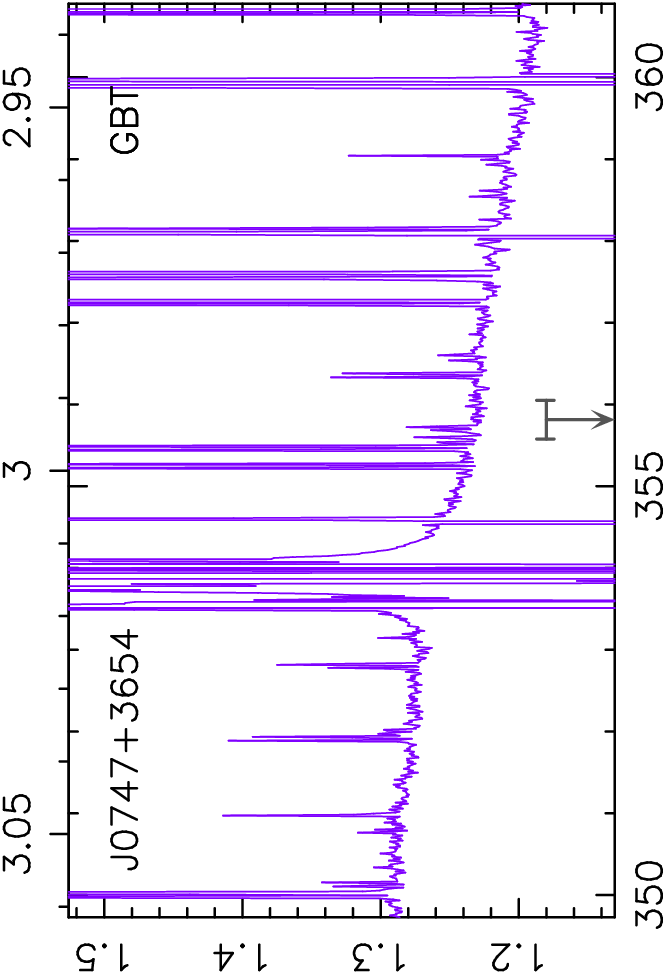}
\includegraphics{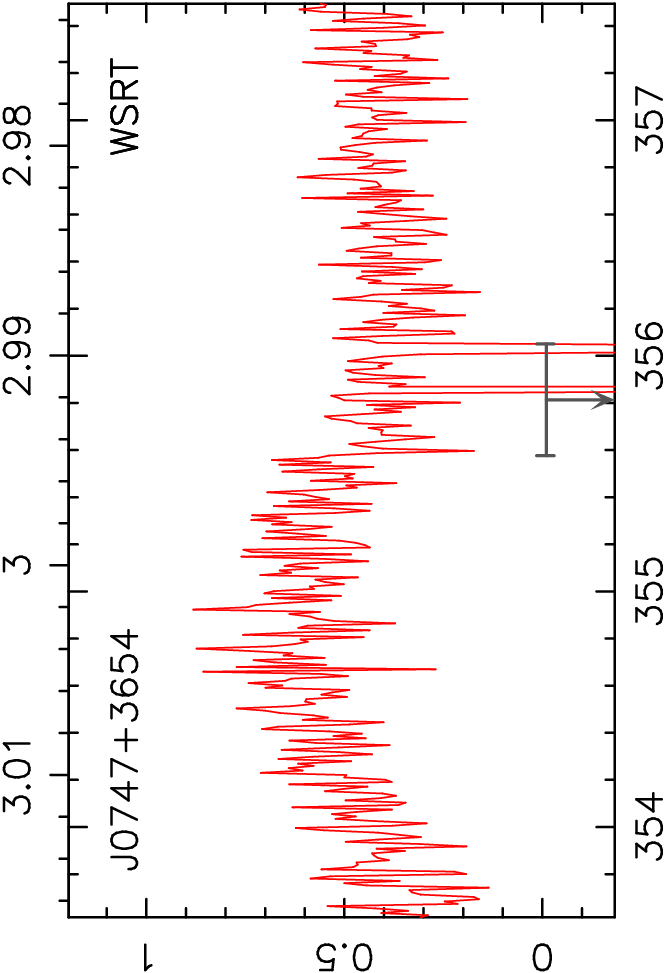}
\includegraphics{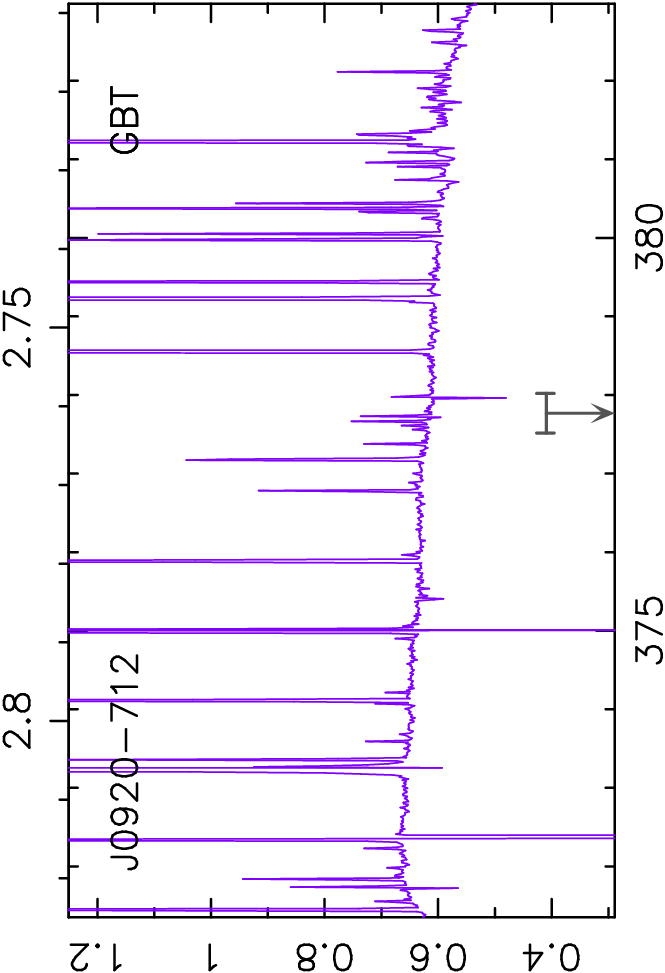}
\includegraphics{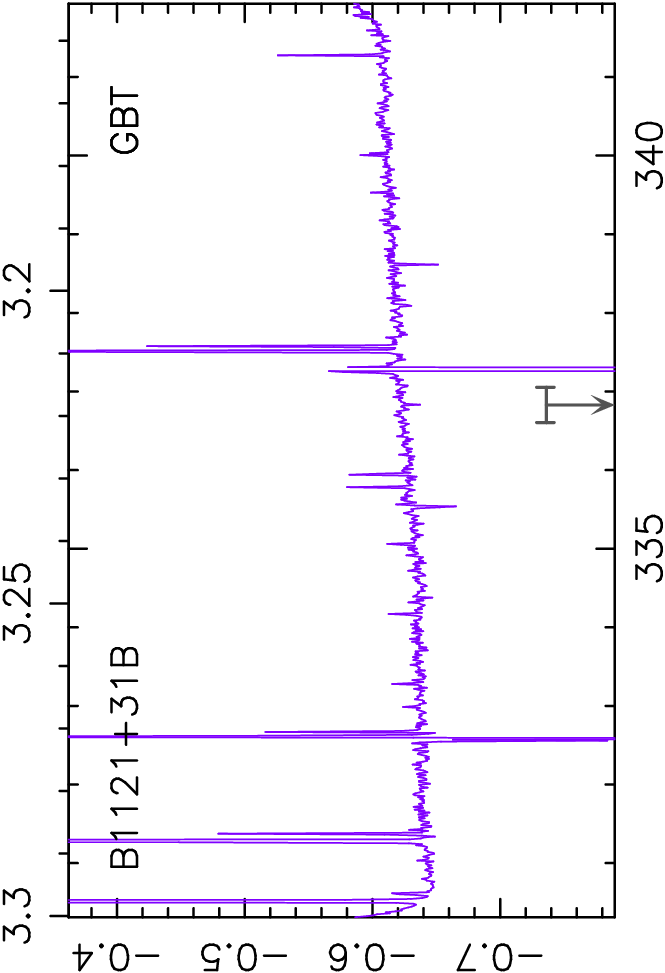}
\includegraphics{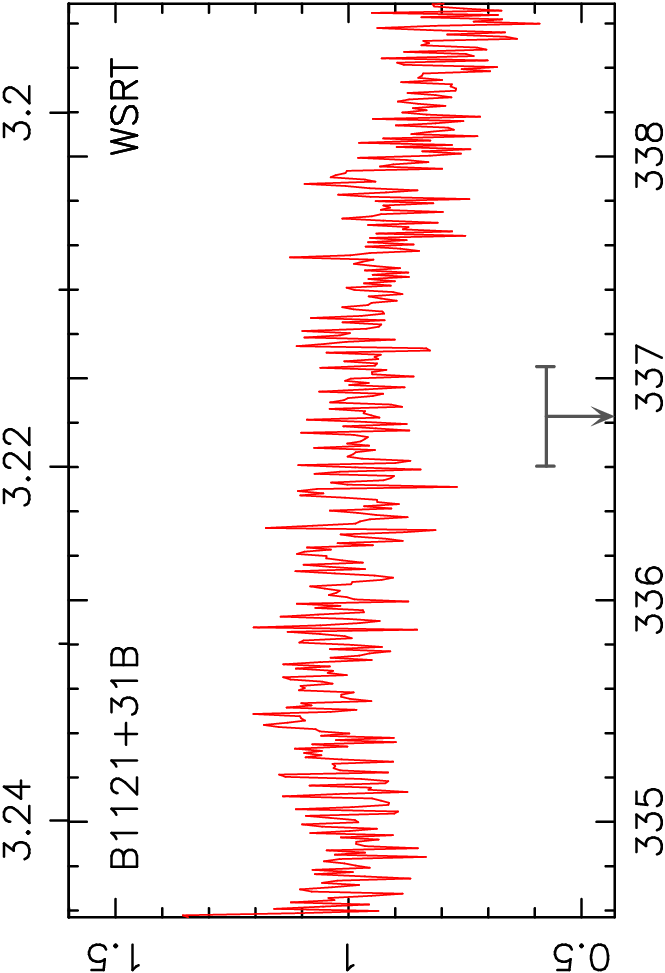}
\includegraphics{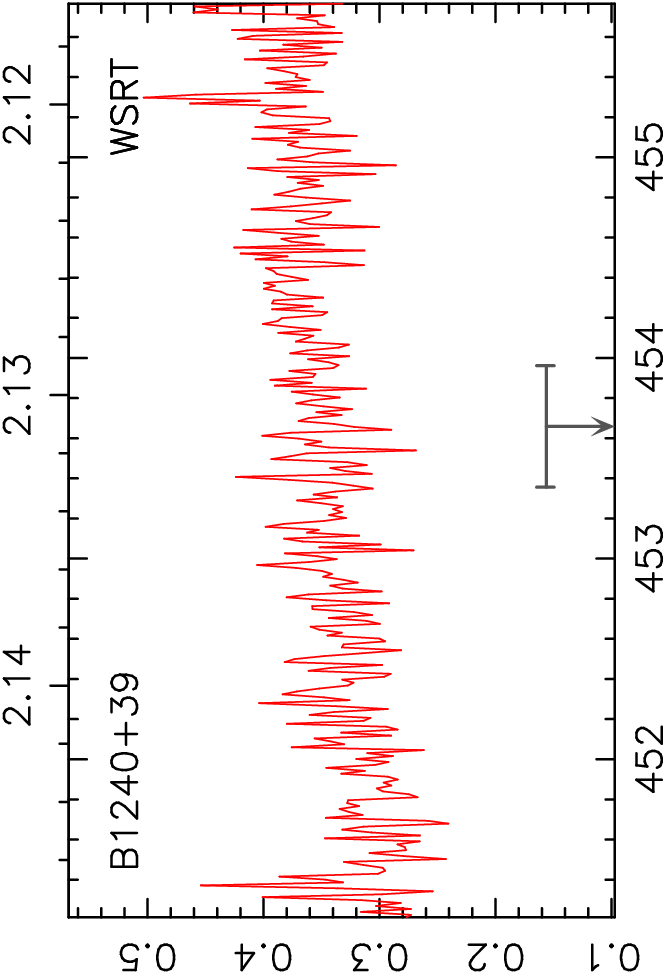}
\caption{The spectra from the \HI\ 21-cm searches shown at a spectral resolution of 10 \kms. The ordinate 
gives the flux density [Jy] and the abscissa the barycentric frequency [MHz]. The downwards arrow shows the expected
frequency of the absorption from the optical redshift, with the horizontal bar showing a span of $\pm200$ \kms\ for guidance. 
The scale along the top axis shows the redshift of \HI\ 21-cm.} 
\label{spectra}
\end{figure*}

For the WSRT observations we requested four hours for each source using all 15 antennas (giving 105 baseline pairs), with the aim of 
reaching an r.m.s. noise level of $\lapp20$ mJy per 10 \kms\ channel. For a flux density of $0.5$ Jy, this gives a $3\sigma$ optical depth limit of
$\tau\approx0.1$ per channel,  a sensitivity to $N_{\rm
  HI}\gapp2\times10^{18}(T_{\rm spin}/f)$ \scm.
We used the UHF-low receiver backed with the correlator over $2\times5$ MHz bands,
each in two polarisations over
1024 channels,  in order to have a channel spacing of $\approx4$ \kms, while
maintaining a redshift coverage of $\Delta z \approx\pm0.04$. For each source 
the  delays  were all self calibrated with 3C\,48, 3C\,147 and 3C\,286
being used for bandpass calibration and the data were reduced using the {\sc
miriad} interferometry reduction package, with a spectrum being extracted from each cube. As per the GBT spectra, these are
shown in Fig. \ref{spectra} and summarised in Table \ref{obs}:\\
{\bf NVSS\,J012142+132058} was observed on 5 January 2010 for a total of 5.0 hours. Severe RFI meant that
extensive flagging of the data was required, leaving 45 baseline pairs. Despite the flagging, a reasonable
image could not be produced and so a spectrum was obtained by averaging together the remaining baselines pairs,
giving a much reduced flux with two spikes remaining. An r.m.s. noise level of 10 mJy per 10 \kms\ channel
was reached between these spikes, which rises to 134 mJy averaged over the whole bandpass. 
Each unsmoothed channel was 4.66 \kms\ wide and the synthesised beam was $772"\times 83"$.\\
{\bf  NVSS\,J020510+224250} was observed on 19 February 2010 for a total of 5.0 hours. Again severe RFI required extensive flagging, leaving only the {\sf XX} polarisation and 78
baseline pairs which produced a poor quality image. A spectrum was, however,  extracted from the cube, which had
an r.m.s. noise level of 77 mJy per 10 \kms\ channel between the two main spikes, which rises to 130 mJy over the whole bandpass.
Each unsmoothed channel was 4.64 \kms\ wide and the synthesised beam $237"\times61"$.\\
{\bf  NVSS\,J023111+360027}  was observed on 10 February 2010 for a total of 3.5 hours. Again, RFI required extensive flagging,
leaving only 55 baseline pairs. From this an image of the target could not be produced, with
patchy emission over the field being evident as a ripple in the extracted spectrum.
Each unsmoothed channel was 4.20 \kms\ wide and the synthesised beam $256"\times69"$.\\
{\bf B2\,0300+37A} was observed for 2.7 hours on 14 December 2009 for 4.5 hours.
After flagging 66 baselines remained, but RFI was still too severe to allow us to produce an image and so we show the
average of the unflagged baseline pairs. Each unsmoothed channel was 3.61 \kms\ wide.\\
{\bf WN\,J0617+5012} was observed at 342.02 MHz
for 6.0 hours on 18 January 2010, but severe RFI meant that no data could be used.\\
{\bf PKS\,0742+10} was observed at 390.30 MHz
for 5.8 hours on 14 December 2009, but again, severe RFI marred the whole observation.\\
{\bf FIRST\,J074729.3+365438} was observed at  355.81 MHz 
for 6.0 hours on 11 March 2010. After flagging the worst RFI, 55 baseline pairs remained from which an image of the
target could not be produced. Nevertheless, the spectrum extracted from the field was relatively spike free and, as in
the case of J0231+3600, there is a strong ripple evident, although the spectrum is very noisy.  Each unsmoothed channel was
4.10 \kms\ wide and the synthesised beam was $244"\times 66"$.\\
{\bf VLSS J1115.1+5016} was observed at 401.24 MHz
on 12 December 2009 for 4.0 hours.
However, severe RFI in this band meant that none of the
data where useable.\\
{\bf B2\,1121+31B} was observed at 336.83 MHz
for 5.2 hours on 19 January 2010.   After flagging, 66 baseline pairs
remained from which an image was produced. The image was of poor quality and the extracted spectrum is over the central $\pm\approx15'$, 
with the flux density showing a strong dependence on the extent of the image used, thus limiting the reliability of this. Each unsmoothed channel was 
4.35 \kms\ wide and the synthesised beam was $288"\times 62"$.\\
{\bf B2\,1240+39} was observed at 453.66 MHz 
for 3.0 hours on 5 April 2010. After flagging, 55 baseline pairs remained and a spectrum was extracted from the cube.
Each unsmoothed channel was 
3.27 \kms\ wide and the synthesised beam was $303"\times 73"$.\\


\begin{table*} 
\centering
\begin{minipage}{160mm}
  \caption{The observational results for the $\gapp0.2$ Jy optically faint northern sources which fall into the GBT and
    WSRT 90-cm bands.  With the exception of B0742+10 (which is from \citealt{wp85}), all are ultra-steep spectrum
    sources (from \citealt{dvs+02}). The second and third columns list the magnitudes and their values. The
    fourth column gives the source redshift, followed by the observed frequency range [MHz] over which $\Delta S$, the
    r.m.s. noise [mJy] reached per 10 \kms\ channel, is applicable. $S$ is the continuum flux density [Jy] and
    $\tau_{\rm obs}=-\ln(1 - 3\Delta S/S_{\rm cont})$ gives the $3\sigma$ limit to the observed optical depth of the line per 10 \kms\
    channel. This is followed by the resulting neutral hydrogen column density, where $N_{\rm HI}=1.823\times10^{18}\,(T_{\rm
      spin}/f)\int\!\tau_{\rm obs}\,dv$) [\scm~K$^{-1}$], $T_{\rm spin}$ being the spin temperature and $f$ the
    covering factor (see Sect. \ref{rp}). Lastly we list the redshift range over which the limit applies.
   }
\begin{tabular}{@{}l l c c c c c  c l c c @{}} 
\hline\smallskip
Source & Mag & Value & $z$  & $\nu_{\rm obs}$  &   $\Delta S$   & $S$ & $\tau_{\rm obs}$ & Tel. &  $N_{\rm HI}.\,(f/T_{\rm spin})$& z-range \\ 
 \hline
0121+1320 & $r_{_{\rm S}}$ & $>24$ &  3.516 &      313.43--316.40    &    23   &  0.455&   $<0.15$  &  GBT  &  $<2.7\times10^{18}$ & 3.4893--3.5318 \\
... &                     ... &                 ... &          ... &               312.11--317.06    &   10   &  0.103  & $<0.29$  & WSRT & $<5.3\times10^{18}$  & 3.4799--3.5510 \\
J0205+2242 & $r_{_{\rm S}}$ & $>24$ &  3.506 &    309.93--317.60  &     24  &     0.251  &   $<0.29$ & GBT &  $<5.3\times10^{18}$ & 3.4723--3.5830\\
... &                     ... &                 ... &          ... &                        367.57--372.16   &    77   & 0.574 & $<0.51$ & WSRT     & $<9.3\times10^{18}$ & 2.8167--2.8643\\
J0231+3600 & $ I $               & $25.0$ &  3.079 &   342.68--353.46   & 189    &  ---   &   ---   &   GBT    & ---  & --- \\ 
... &                     ... &                 ... &          ... &                    345.80--350.75  & 5.3 & 0.377 &$<0.042$ & WSRT & $<7.7\times10^{17}$ & 3.0496--3.1076 \\
B0300+37A& $r_{_{\rm S}}$  & $23.2$ &  2.506 & 400.94--408.65  &      ---          &    ---      &      ---      &   WSRT    & ---  & ---  \\ 
J0617+5012 & $R$              & $ > 24 $&  3.153 &     337.56 --343.17  &    17          &    1.277   &   $<0.040$  &   GBT    & $<7.3\times10^{17}$ &   3.1391--3.2079\\
... &                     ... &                 ... &          ... &            339.59-- 344.53       &    ---      &      ---      &      ---    &   WSRT    &  ---  & --- \\ 
B0742+10  &   $R $             & $23.7$ & 2.630  &  391.32--394.74    &    22      &     0.874    &  $<0.076$    &  GBT  & $<1.4\times10^{18}$ &  2.5983--2.6298\\
.. &                     ... &                 ... &          ... &              388.78--393.61 &    ---      &      ---      &   ---   & WSRT &  ---  & ---   \\ 
J0747+3654 & $r_{_{\rm S}}$ & $ >23$ & 2.992 &   355.78--357.15  &  15  &    1.232   &    $<0.037$  &   GBT   & $<6.7\times10^{17}$ &     2.9771--2.9923  \\
.. &                     ... &                 ... &          ... &            353.39--358.34 & 100 & 0.466 &  $<1.0$ & WSRT & $<1.8\times10^{19}$ &   2.9638--3.0194\\
J0920--0712 & $R$              & $22.4$  & 2.760  &  --- & --- & 0.621      & ---    &  GBT   & ---  & ---   \\
J1115+5016 &  $r_{_{\rm S}}$& $  >24$ & 2.540  &  396.21--406.21  &    ---             &    ---                  &    ---                &     WSRT    & ---  & --- \\
B1121+31B & $r_{_{\rm S}}$ & $21.3$ &  3.217 &   332.74--337.16  &  7.9&  ---& $<0.042^{\ast}$ &  GBT   & $<7.7\times10^{17}$ &   3.2129--3.2688 \\ 
. &                     ... &                 ... &          ... &                 334.56-- 338.70  & 66 & 0.876 & $<0.23$ & WSRT& $<4.2\times10^{18}$ & 3.1937--3.2456\\
B1240+39 & $r_{_{\rm S}}$ & $23.6$ & 2.131  & 451.17--456.15  & 39  &  0.348  & $<0.41$  & WSRT & $<7.5\times10^{18}$ & 2.1139--2.1483  \\ 
\hline 
\end{tabular}
{Notes: $^{\ast}$Derived by interpolating $S_{\rm 151~MHz} = 1.19$ Jy \citep{wyrw96} and $S_{\rm 365~MHz} = 0.525$ Jy \citep{dbb+96}, giving  a continuum flux of $S_{\rm 378~MHz} \approx 0.57$ Jy.}
\label{obs}
\end{minipage}
\end{table*}

\section{Possible reasons for the non-detection of 21-cm}
\subsection{Sensitivity limits}
\label{odl}

Despite our deliberate selection of optically faint sources, with continuum luminosities below the threshold of
$L_{1216}=10^{23}$ \WpHz, we have not detected \HI\ 21-cm in any of the eight sources for which there were useable data.
 Below this critical value, \citet{cw10} find a $\gapp40$\% general detection rate, which consists of $\approx50$\%
  for compact objects\footnote{Gigahertz peaked spectrum (GPS), compact steep spectrum sources (CSS) and
    compact symmetric objects (CSO). Although the 21-cm detection rate is generally believed to be higher for these
than the general population, when the $L_{1216}\geq10^{23}$ \WpHz\ sources are removed the rates
are similar, indicating that the higher rates may be due to the generally
low ultra-violet luminosities of the compact objects \citep{cw10}.} and $\approx40$\% for others. That is, we may expect approximately three detections out 
of the eight sources for which the data  were not overwhelmed by RFI.

\begin{figure*}
\centering \includegraphics[angle=-90,scale=0.45]{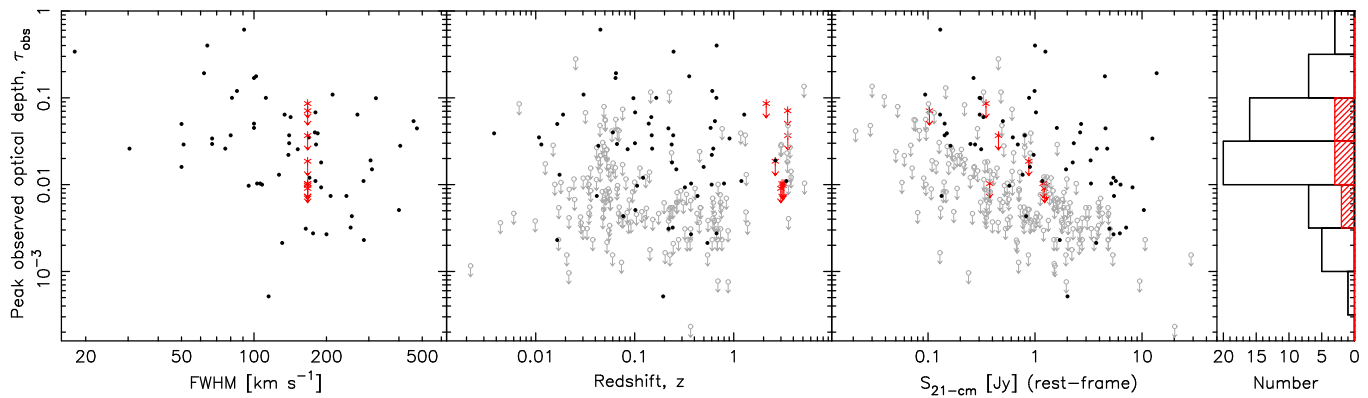}
\caption{The observed optical depth versus the FWHM of the profile (left), redshift (middle) and rest-frame 1.4 GHz flux density (right).
 The symbols are as per Fig. \ref{M-z}, where in the left panel, for the sake of clarity, only our limits are shown.
These, and the other limits in the remaining panels, have been resampled to a spectral resolution of 167 \kms, the mean of the detections which range from 18 to 475 \kms\ 
(\citealt{cps92} and \citealt{vpt+03}, respectively). The unfilled histogram shows the distribution for the detections and the hatched histogram the limits to our
eight targets not overwhelmed by RFI.}
\label{tau_dist}
\end{figure*}

One possible reason for the non-detections is that all of the sources where affected to some degree by RFI, giving some 
relatively poor sensitivity limits (Table \ref{obs}), where detections of optical depths of $\tau_{\rm obs}\gapp0.1$ have been documented
to be in the minority \citep{vpt+03,mor04}. In order to investigate this, in Fig. \ref{tau_dist} we show the obs optical depths
for all the redshifted 21-cm searches.
Since many of our  sources were observed by both the GBT and WSRT, we use the best limit in each case and
smooth these, as well as rebinning the $3\sigma$ limits from the literature, to a channel spacing of 167 \kms, the
mean FWHM of the 21-cm detections. This normalises all of the limits, published at various spectral resolutions,
and gives the limit for the detection of putative line of width FWHM$\,=167$ \kms\ within a single channel.

This normalisation has the effect of moving all of our limits to $\tau_{\rm obs} \lapp 0.1$, while
degrading the column density limits (see Fig.~\ref{3-N}) by a factor of $\approx4$ to those quoted in Table \ref{obs}.
From Fig. \ref{tau_dist}, we see that our limits are  located in the densest clustering of  the detections and so
for the rest of the analysis, since we are attempting
to explain the exclusive non-detections, we assume that (at least some of) the targets have been searched sufficiently deeply.

We note also a weak correlation between the obs optical depth and the profile width (left panel)\footnote{The observed
distribution has a 1.4\% probability of occuring by chance, which is significant at $2.46\sigma$, assuming Gaussian
statistics.}, which may suggest that weak, wide profiles may be missed by 21-cm surveys, despite having 
similar column densities to the detections (Allison et al., in prep.). 
For completeness, we also show how the optical depth is distributed with redshift (middle panel) and
the rest frame 1.4~GHz flux density, from which we see the expected anti-correlation.


\subsection{Radio properties}
\label{rp}

A possible effect which could reduce the probability of detecting 21-cm absorption is the non-deliberate selection of
these high redshift radio galaxies having steep radio spectra (all, apart from B0742+10 being ``Ultra-Steep Spectrum
Radio Sources'', \citealt{dvs+02}).  This may have an effect on the {\em actual} optical depth of the line, $\tau$,
which is related to the {\em observed} optical depth, $\tau_{\rm obs}\equiv\Delta S/S$, through
$\tau\equiv-\ln\left(1-\frac{\Delta S}{Sf}\right)$, where $f$, the covering factor, quantifies how much of the observed
flux is intercepted by the absorber. In the optically thin regime (where $\Delta S \lapp 0.3\,S$), this expression
reduces to $\tau\approx {\Delta S/}({Sf}) = \tau_{\rm obs}/f$. Since, by definition, $f\leq1$ then the actual optical
depth limit is $\tau\geq\tau_{\rm obs}$.\footnote{Since $f$ is generally unknown (see \citealt{cur12}), we leave this
  (as well as the spin temperature) as a free parameter (Table \ref{obs}). Assuming $f$ has its maximum value of unity,
  as if often the case in the literature, improves the perceived sensitivity of the observation, where in fact $f<1$.}
That is, the limit is affected by how effectively the absorbing gas intercepts the background emission.

 Unfortunately, for this sample there is only high
resolution radio imaging available for one source -- B0742+10, which subtends 1.2 mas at 15 GHz \citep{sdo+01}. At
$z = 2.630$ this corresponds to a linear extent of 10 pc (using a
standard $\Lambda$ cosmology with $H_{0}=71$~km~s$^{-1}$~Mpc$^{-1}$, $\Omega_{\rm matter}=0.27$ and
$\Omega_{\Lambda}=0.73$). Of the remainder of the sample, only 
four are in the area of sky currently covered by the Very
Large Array's FIRST (Faint Images of the Radio Sky at Twenty
Centimetres) survey, having deconvolved minor axes of 0.91\arcsec (7 kpc) for J0747+3654, 1.02 \arcsec (8 kpc) for J0920--0712,
1.58\arcsec (12 kpc) for B1121+31B and 0.75\arcsec (6 kpc) for B1240+39. Since these are typically unresolved by the 5\arcsec
synthesised beam, assigning a source size of $\lapp40$ kpc for each component may be more apt.
For J0121+1320, J0205+2242, J0231+3600, B0300+37A and
J0617+5012, which are all unresolved by the NRAO VLA Sky Survey (NVSS), the typical 17\arcsec minor axis of the beam gives
source sizes of $\lapp130$ kpc for these redshifts. 

\begin{figure*}
\centering \includegraphics[angle=-90,scale=0.60]{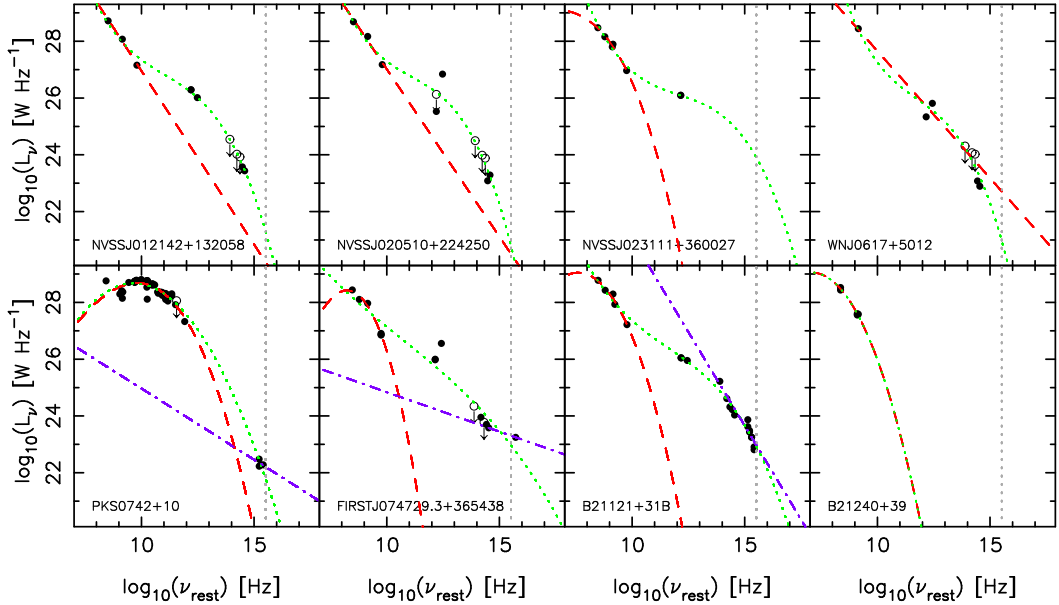}
\caption{The SEDs of our eight targets not overwhelmed by RFI. The broken curve shows
the fit to the radio data, the dotted line to the full data and the broken/dotted to the UV data, where possible (see the Appendix).
The vertical dotted line signifies a rest-frame frequency of $3.29\times10^{15}$ Hz ($\lambda = 912$ \AA).}
\label{SEDs}
\end{figure*}

Given the lack of high resolution radio imaging, information on the source size may be gleaned from the radio SEDs, given that
the turnover frequency is anti-correlated with the size of
the radio source \citep{ffs+90}.
In order to explore this possibility, in Fig.~\ref{SEDs}, we show the spectral energy distributions (SEDs) of our eight targets,
from which the derived turnover frequencies and spectral indices are summarised in Table \ref{lums}.
\begin{table} 
\centering
\begin{minipage}{75mm}
\caption{The radio properties and luminosities of the targets. $\nu_{_{\rm TO}}$ is the
  rest-frame turnover frequency [GHz], where the upper limits designate no observed turnover (which is thus assumed to
  occur below the observed frequencies), followed by the spectral index at the rest-frame 21-cm frequency, \AL. In the last
  three  columns we list the $\lambda=1216$ \AA\ (as originally estimated and not corrected for extinction) and $\lambda=912$ \AA\ 
(as derived from the fits in the Appendix) rest-frame continuum luminosities [\WpHz], as well as
  the ionising ($\lambda\leq912$ \AA) photon rate [s$^{-1}$] in logs.}
\begin{tabular}{@{}l r r  c c c   @{}} 
\hline \smallskip
Source                & $\nu_{_{\rm TO}}$ & $\alpha$ & $L_{1216}$ & $L_{912}$ & rate \\
\hline
J0121+1320 &  $<0.331$ & $-1.22$  &   22.50&  --- &  ---  \\
J0205+2242 &  $<0.331$ &   $-1.18$&22.15& ---  &  --- \\
J0231+360 &  $<0.302$ &  $-1.18$  & 21.28 &---   & ---   \\
J0617+5012 & $<1.41$ &  $-0.89$ & 22.00 & ---  &---   \\
B0742+10 & 6.46 & $0.29$  & 22.00  & 22.20  & 55.68   \\
J0747+3654 &  0.17&$-1.23$   & 22.54 & 23.30  & 57.04  \\
B1121+31B &  0.040& $-1.21$  & 22.98 & 22.97  & 56.03  \\
B1240+39 &  0.012&  $-1.50$   &  22.60& ---  & ----  \\
\hline       
\end{tabular}
 \label{lums}  
\end{minipage}
\end{table} 
From this we see that B0742+10 is the only source with an indisputable turnover, consistent with its observed
compactness (10 pc at 15 GHz, \citealt{sdo+01}). Given that none of the other sources exhibit a {\em clear} turnover
above the lowest frequency which has been observed in the literature\footnote{Whether there are  turnovers in the SEDs of J0231+360,
  J0747+3654, B1121+31B and B1240+39 are disputable (Fig. \ref{SEDs}).}, it is possible that the majority of our targets may have very
extended radio morphologies, thus making $\tau\gg\tau_{\rm obs}$ in the case of the limits.

In order to verify this through the comparison of the radio SEDs with the rest of the 21-cm searched sources (compiled in \citealt{cw10,ace+12}),
  we fit the radio SEDs (see the Appendix) and use the {\sc asurv} survival analysis package \citep{ifn86} to obtain
  mean values of the turnover frequency and spectral index for various sub-samples (Table \ref{stats}).
\begin{table} 
\centering
\begin{minipage}{82mm}
  \caption{The mean values of the turnover frequency [Hz] and the spectral index at the rest-frame 1420 MHz for all of
    the sources searched in redshifted 21-cm absorption (\citealt{cw10,ace+12}). $n_{\rm p}$ gives the number of data points and
 $n_{\rm l}$ the number of these which are limits.  Note that the various sub-samples from the second row on
  exclude the current sources, which are summarised in the first row.}
\begin{tabular}{@{} l c r r  c r @{}}
\hline\smallskip
Sub-sample               & \multicolumn{3}{c}{\sc turnover frequency} & \multicolumn{2}{c}{\sc spectral index}\\
                           & $\left<\log_{10}\nu_{_{\rm TO}}\right>$ &  $n_{\rm p}$ &  $n_{\rm l}$&   $\left<\alpha\right>$ & $n_{\rm p}$ \\
\hline
This paper                                 &  $7.91\pm0.35$ & 8  & 4     &  $-1.02\pm 0.18$ & 8 \\
Detections                    & $8.19\pm0.14$ &  56  & 23 &    $-0.29\pm 0.08$ & 58\\
Non-detections                        &  $8.03\pm0.09$ & 171 & 92 &  $-0.39\pm0.04$ & 173 \\
Dets + nons                             & $8.07\pm0.08$ & 227 & 115 & $-0.37\pm 0.04$ & 231\\
UV luminous                             &  $8.14\pm0.29$ & 19    & 11  & $-0.26\pm0.12$ & 19\\
\hline       
\end{tabular}
\label{stats}
\end{minipage}  
\end{table}
From this it is seen that the mean turnover frequency for this sample is indeed lower than for any of
the other sub-samples, indicating that the radio sources are generally larger. However, this is based upon four limits to the turnover frequencies 
out of eight SEDs and, as warned by {\sc asurv},
may be unreliable.\footnote{As seen from Fig. \ref{SEDs},
it is conceivable that seven of the turnover frequencies are limits. Furthermore, the high redshift sources will also be subject to a higher mean turnover due to that
  fact that the lowest typical value, 74 MHz, is in the observed-frame, giving the high redshift sources higher upper
  limits in the rest-frame.} In the case of the spectral indices, where there are no limits, we see that the spectra of our sample are
significantly steeper than for the other sub-samples.\footnote{Note also that the non-detections have steeper indices
  than the detections, although these are consistent within the uncertainties.} Thus, if the spectral index is a
reliable tracer of the background source size, the fact that this sample consists almost exclusively of steep spectrum sources, indicating
large radio source sizes, means that poor coverage of the background source cannot be ruled out as the cause of the
non-detections.  Note, however, for the ``UV luminous'' sources (the 19 which lie above the $L_{\rm UV}=10^{23}$ \WpHz\
cut-off, \citealt{cwm+10}), the mean spectral index is close to that of the 21-cm detections, thus ruling out the same
predisposition towards large radio sources being the cause of the non-detections. 

\subsection{Ionising luminosities and photon rates}
\label{ilpr}

Although all of our sources were selected to have $\lambda=1216$ \AA\  continuum 
 luminosities below the critical value of $L_{\rm 1216}\sim10^{23}$ \WpHz\ (Fig. \ref{M-z}), it is clear that
the total ionising ($\lambda\leq912$ \AA) luminosity, $\int^{\infty}_{\nu_{\rm ion}}({L_{\nu}}/{\nu})\,d{\nu}$, provides a much better measure of the ionising flux
than a monochromatic value (be it $L_{\rm 1216}$ or $L_{\rm 912}$, \citealt{cw12}). In equilibrium, the total ionising luminosity
is related to the recombination of the atoms via \citep{ost89},
\begin{eqnarray}
  \int^{\infty}_{\nu_{\rm ion}}\frac{L_{\nu}}{h\nu}\,d{\nu}= 4\pi\int^{r_{\rm ion}}_{0}\,n_{\rm p}\,n_{\rm e}\,\alpha_{A}\,r^2\, dr ,
\label{eq1}
\end{eqnarray}
where $h$ is the Planck constant, giving the number of ionising photons per second. On the right hand side of the
expression, $n_{\rm p}$ and $n_{\rm e}$ are the proton and electron densities, respectively, $\alpha_{A}$ the radiative
recombination rate coefficient of hydrogen and $r_{\rm ion}$ is the extent of the ionisation (the ``Str\"{o}mgren
sphere'').  For a gas which decreases in density with increasing
distance from the ionising source,
\citet{cw12} show that $r_{\rm ion}\rightarrow\infty$ for a finite ultra-violet luminosity and that for a large spiral this value is $L_{912} \sim L_{1216} \sim 10^{23}$ \WpHz, thus
explaining this critical luminosity above which 21-cm absorption has never been detected.

\begin{figure*}
\centering \includegraphics[angle=0,scale=0.70]{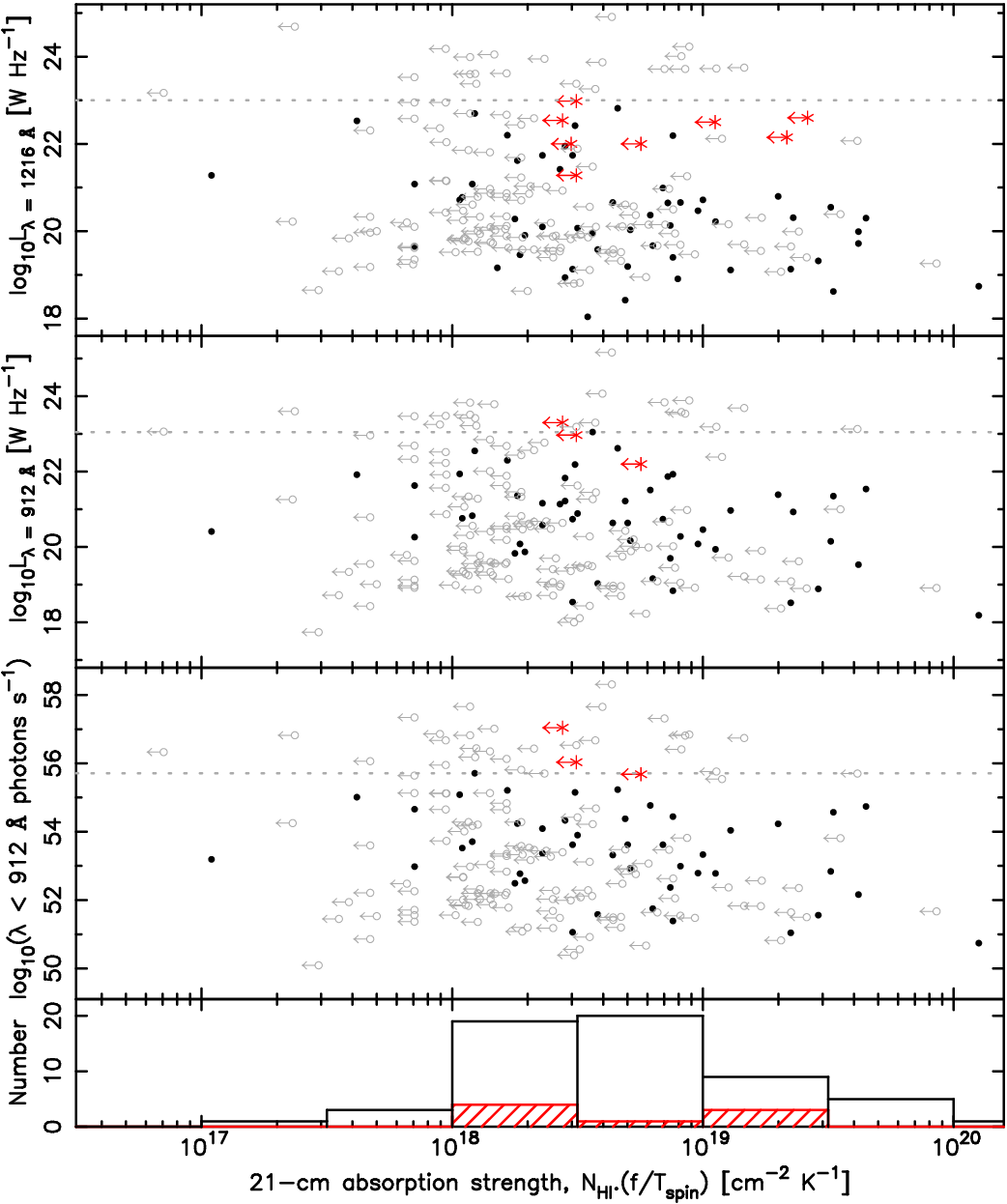}
\caption{The $\lambda=1216$ \AA\ continuum luminosity (top), $\lambda=912$ \AA\ continuum luminosity (middle) and number 
of ionising photons per second (bottom) versus the 21-cm line strength ($1.823\times10^{18}\int\tau dv$), see Sect. \ref{ilpr}. 
The horizontal lines in the top and middle panels shows the critical monochromatic luminosity of $10^{23}$ \WpHz\ and
in the bottom panel $\int^{\infty}_{\nu_{\rm ion}}({L_{\nu}}/{h\nu})\,d{\nu} = 5.1\times10^{55}$ s$^{-1}$, the highest photon
rate  at which a 21-cm detection occurs (see main text). The symbols are as per Fig. \ref{M-z},
with the left pointing arrows showing the $3\sigma$ upper limits to the 21-cm line strength (from \citealt{cw10,cwm+10,ace+12}).
As per Fig. \ref{tau_dist}, these have been rebinned and integrated over a FWHM$\,=167$ \kms, which has the effect of
degrading our limits by a factor of $\approx4$ in comparison to those quoted in Table \ref{obs} (see \citealt{cur12} for details).
The unfilled histogram shows the distribution of the detection line strengths and the hatched histogram the limits to
eight targets not overwhelmed by RFI.}
 \label{3-N}
\end{figure*}

From the mean SEDs of all of the associated 21-cm searches, the critical ionising photon rate arising from this luminosity is estimated to be
$\int^{\infty}_{\nu_{\rm ion}}({L_{\nu}}/{h\nu})\,d{\nu} = 2.9\times10^{56}$ s$^{-1}$ \citep{cw12}.
From the polynomial fits (see the Appendix), we could determine photon rates for three of the targets presented here 
(Fig.~\ref{SEDs}), one of which is above the estimate of the critical value 
($\log_{10}[2.9\times10^{56}] =56.5$, cf. Table~\ref{lums}). Applying fits to each of the SEDs,
the highest photon rate which could be determined for a 21-cm detection is $5.1\times10^{55}$ s$^{-1}$,
which is below or close to the photon rate of the three targets  ($\log_{10}[5.1\times10^{55}] =  55.7$, see Fig.~\ref{3-N}).
However, given that we have no estimates of the photon rates for the remaining five sources, 
 from the previous estimates of the $\lambda = 1216$~\AA\
continuum luminosities (Table~\ref{lums}), it is possible that several of these have rates lower than for the three for
which we do have estimates. For instance, J0121+1320 has been detected in CO emission \citep{dno03}, which
may suggest a UV luminosity below the critical value, although further studies are required to find if this applies to 
the warm emitting molecular gas.
While the monochromatic luminosities may not provide a reliable
estimate of the photon rates, if, for the sake of argument, the remaining five sources do have ionising photon rates below the critical value,
for a 40\% detection rate (Sect. \ref{odl}), the binomial probability of zero out of five detections is 0.078. This is 
significant at $1.76\sigma$, assuming Gaussian statistics, and so not statistically important. Recalling that a possible reason for the 
non-detections is that the targets have not been searched sufficiently deeply (Sect. \ref{odl}), this significance could be lower, thus
not requiring us to ``explain away'' all eight non-detections.

Lastly, given that $\int^{\infty}_{\nu_{\rm ion}}({L_{\nu}}/{h\nu})\,d{\nu} \approx3\times10^{56}$ s$^{-1}$ is sufficient to ionise all of the gas
in a large spiral, photon rates lower than this will completely ionise gas disks of correspondingly smaller
scale-lengths \citep{cw12}. The sample presented here is at more than double the look-back time of the majority
of 21-cm detections (Fig.~\ref{M-z}) and so if there 
is any evolution in galactic morphology, with a larger fraction of smaller galaxies at
higher redshift \citep{bmce00,lf03}, we may expect the value of the critical photon rate to be lower at the redshifts searched here, making
the detection of 21-cm absorption all the more difficult. 

Evolution may also play a r{\^o}le in that if the molecular (\MOLH) component constitutes a larger fraction of the
  gas at high redshift \citep{or09a}, lower relative column densities will make the \HI\ more difficult to
  detect. However, as seen from Fig. \ref{M-z}, 21-cm has already been detected at $z\gapp2$
  and, although molecular gas has been detected in emission in over 100 ultra-luminous
  infrared galaxies (see \citealt{cur09,ctk+11} and references therein), searches for molecular absorption within the hosts of 
  millimetre-loud sources have proved unfruitful \citep{cwc+11}.\footnote{Given that J0121+1320 is detected in CO emission
    and that a further five of the targets are known exhibit strong FIR emission (Fig.~\ref{SEDs}), 
    these sources should be searched in the millimetre band to rule out the
    possibility of a high molecular fraction suppressing the atomic abundance (if not already done so, although there are no
    published searches for molecular emission in the remaining targets). However, the cool absorbing molecular gas,
    rather than warm emitting component, may be expected to be the coincident with the cool atomic gas (see
    \citealt{cw98b,mwf+00} and references therein).}

\section{Summary}

We have undertaken a survey for the hydrogen apparently missing in high redshift radio
galaxies and quasars by selecting targets in which
the optical magnitudes indicate that these are below the UV luminosity cut-off, above which all of the
gas is ionised \citep{cw12}. Despite this, there were no detections in the eight $z\gapp2$ objects 
for which useful data were obtained. Upon an examination of the spectral energy distributions of the targets,
we suggest two possible reasons for this:
\begin{enumerate}
\item Due to our requirement of optically faint, radio-loud objects, for which their redshift placed them within the
  GBT and WSRT 90-cm bands, our sample was dominated by the ultra-steep spectrum sources of \citet{dvs+02}. By fitting
  radio SEDs to these, we find a mean spectral index of $\left<\alpha\right> = -1.0$, which is significantly higher than
  $\left<\alpha\right> = -0.3$ and $-0.4$ for the 21-cm detections and the remaining  non-detections, respectively. This
  suggests that most of our targets may have extended radio emission, thus reducing the coverage of the background flux
  by the absorbing gas, lowering the effective sensitivity of our survey. 

    \item Although our magnitude selection predicted that all of the targets are below the critical luminosity of $L_{\rm UV}\sim10^{23}$ \WpHz,
      those for which the ionising photon rate could be determined are above or close to $5.1\times10^{55}$ ionising photons s$^{-1}$, the
      highest value which can be determined for a 21-cm detection. However, the photon rate could only be estimated for three of the
      eight sources and their monochromatic $\lambda=1216$ \AA\  continuum  luminosities may nevertheless suggest that
      these are below the threshold. Without sufficient blue/ultra-violet photometry, this is, however, speculation.
\end{enumerate}
Although it is impossible at this time to determine which reason is main the culprit, note that the mean radio-band SEDs of the 19
sources for which $L_{1216} \gapp10^{23}$ \WpHz, which initially highlighted a UV threshold, is no steeper than the SEDs of the 21-cm detections. This
suggests that a bias towards larger radio sources is not the cause of the exclusive
non-detections for the UV luminous sample.
Also, presuming that the remaining five sources, for which the photon rate could not be determined, are below the
threshold gives a detection rate (0 out of 5) which is not statistically significant, given the 40\% chance of a detection
at $z\lapp1$.

Furthermore, at these redshifts we are probing look-back times more than double that of the $z\lapp1$ sources and if
there is a larger fraction of smaller galaxies at these epochs, the gas will be ionised at lower luminosities.  For
instance, the highest photon rate for which there exists a 21-cm detection ($5\times10^{55}$ s$^{-1}$), is sufficient to
fully ionise a gas distribution of scale-length of 1.6 kpc (50\%  that of a large spiral, \citealt{kk09}), for $n_0 = 10$ \ccm\ \citep{cw12}.\footnote{For a disk, in
  which the thickness flares in the same manner as for the Milky Way \citep{kdkh07}, the total gas mass is 
estimated to be $M_{\rm gas}
  =1.3\times10^{9}$ \Mo, a sixth of that  found for a rate of $3\times10^{56}$ ionising photons s$^{-1}$, the critical
  value estimated by \citet{cw12}.} If this is the case, objects even fainter than those targetted here are required to
find the missing hydrogen at high redshift, a task perhaps best suited for blind surveys with the Square Kilometre Array, where
an optical redshift is not a prerequisite.

To  conclude, although we cannot rule out the radio structure and other hitherto unforeseen high redshift effects as the
source of the non-detections, the fact remains that $\lambda\leq912$ \AA\ photons ionise hydrogen and there exists
a $8.32\times10^{-8}$ probability (a $5.36\sigma$ significance) of the exclusive non-detections above a given ionising photon
rate arising by chance \citep{cw12}. This is conjunction with the fact that the critical rate observed is just sufficient to ionise all of the gas in a large
galaxy, leaves little doubt that photoionsation of the gas by the active nucleus is a dominant issue in the search of neutral
gas within the hosts of these objects.


\section*{Acknowledgements}

We wish to thank Gyula J\'{o}zsa for coordinating all of the WSRT observations.
This research has made use of the NASA/IPAC Extragalactic Database
(NED) which is operated by the Jet Propulsion Laboratory, California
Institute of Technology, under contract with the National Aeronautics
and Space Administration. This research has also made use of NASA's
Astrophysics Data System Bibliographic Services and {\sc asurv} Rev
1.2 \citep{lif92a}, which implements the methods presented
in \citet{ifn86}.
The Centre for All-sky Astrophysics is an Australian Research Council Centre of Excellence, funded by grant CE110001020.


\begin{thebibliography}{}

\bibitem[\protect\citeauthoryear{{Allison}, {Curran}, {Emonts}, {Gereb},
  {Mahoney}, {Reeves}, {Sadler}, {Tanna} \& {Whiting}}{{Allison}
  et~al.}{2012}]{ace+12}
{Allison} J.~R., et al.,  2012,
  MNRAS, 423, 2601

\bibitem[\protect\citeauthoryear{{Baker}, {Mathlin}, {Churches} \&
  {Edmunds}}{{Baker} et~al.}{2000}]{bmce00}
{Baker} A.~C.,  {Mathlin} G.~P.,  {Churches} D.~K.,    {Edmunds} M.~G.,  2000,
  in Favata F.,  Kaas A.,   Wilson A.,  eds, Star Formation from the Small to
  the Large Scale, Vol.45 of ESA SP ESA Special Publication, {The Chemical
  Evolution of the Universe}.
Noordwijk, p.~21

\bibitem[\protect\citeauthoryear{{Carilli}, {Perlman} \& {Stocke}}{{Carilli}
  et~al.}{1992}]{cps92}
{Carilli} C.~L.,  {Perlman} E.~S.,    {Stocke} J.~T.,  1992, ApJ, 400, L13

\bibitem[\protect\citeauthoryear{{Combes} \& {Wiklind}}{{Combes} \&
  {Wiklind}}{1998}]{cw98b}
{Combes} F.,  {Wiklind} T.,  1998, ESO Messenger, 91, 29

\bibitem[\protect\citeauthoryear{Curran}{Curran}{2009}]{cur09}
Curran S.~J.,  2009, A\&A, 497, 351

\bibitem[\protect\citeauthoryear{Curran}{Curran}{2012}]{cur12}
Curran S.~J.,  2012, ApJ, 748, L18

\bibitem[\protect\citeauthoryear{Curran, Tanna, Koch, Berengut, Webb, Stark \&
  Flambaum}{Curran et~al.}{2011a}]{ctk+11}
Curran S.~J.,  Tanna A.,  Koch F.~E.,  Berengut J.~C.,  Webb J.~K.,  Stark
  A.~A.,    Flambaum V.~V.,  2011a, A\&A

\bibitem[\protect\citeauthoryear{Curran \& Webb}{Curran \& Webb}{2006}]{cw06}
Curran S.~J.,  Webb J.~K.,  2006, MNRAS, 371, 356

\bibitem[\protect\citeauthoryear{Curran \& Whiting}{Curran \&
  Whiting}{2010}]{cw10}
Curran S.~J.,  Whiting M.~T.,  2010, ApJ, 712, 303

\bibitem[\protect\citeauthoryear{Curran \& Whiting}{Curran \&
  Whiting}{2012}]{cw12}
Curran S.~J.,  Whiting M.~T.,  2012, ApJ, in press (arXiv:1204.2881)

\bibitem[\protect\citeauthoryear{Curran, Whiting, Combes, Kuno, Francis, Nakai,
  Webb, Murphy \& Wiklind}{Curran et~al.}{2011b}]{cwc+11}
Curran S.~J.,  et al.,  2011b, MNRAS, 416, 2143

\bibitem[\protect\citeauthoryear{Curran, Whiting, Murphy, Webb, Bignell,
  Polatidis, Wiklind, Francis \& Langston}{Curran et~al.}{2011c}]{cwm+10}
Curran S.~J.,  et al.,  2011c, MNRAS, 413,
  1165

\bibitem[\protect\citeauthoryear{Curran, Whiting \& Webb}{Curran
  et~al.}{2009}]{cww09}
Curran S.~J.,  Whiting M.~T.,    Webb J.~K.,  2009, Proceedings of Science, 89,
  Chap. 11

\bibitem[\protect\citeauthoryear{Curran, Whiting, Webb \& Athreya}{Curran
  et~al.}{2011d}]{cwwa11}
Curran S.~J.,  Whiting M.~T.,  Webb J.~K.,    Athreya A.,  2011d, MNRAS, 414,
  L26

\bibitem[\protect\citeauthoryear{Curran, Whiting, Wiklind, Webb, Murphy \&
  Purcell}{Curran et~al.}{2008}]{cww+08}
Curran S.~J.,  Whiting M.~T.,  Wiklind T.,  Webb J.~K.,  Murphy M.~T.,
  Purcell C.~R.,  2008, MNRAS, 391, 765

\bibitem[\protect\citeauthoryear{{De Breuck}, {Neri} \& {Omont}}{{De Breuck}
  et~al.}{2003}]{dno03}
{De Breuck} C.,  {Neri} R.,    {Omont} A.,  2003, New Astronomy Review, 47, 285

\bibitem[\protect\citeauthoryear{{De Breuck}, {van Breugel}, {Stanford},
  {R{\"o}ttgering}, {Miley} \& {Stern}}{{De Breuck} et~al.}{2002}]{dvs+02}
{De Breuck} C.,  {van Breugel} W.,  {Stanford} S.~A.,  {R{\"o}ttgering} H.,
  {Miley} G.,    {Stern} D.,  2002, AJ, 123, 637

\bibitem[\protect\citeauthoryear{{Douglas}, {Bash}, {Bozyan}, {Torrence} \&
  {Wolfe}}{{Douglas} et~al.}{1996}]{dbb+96}
{Douglas} J.~N.,  {Bash} F.~N.,  {Bozyan} F.~A.,  {Torrence} G.~W.,    {Wolfe}
  C.,  1996, AJ, 111, 1945

\bibitem[\protect\citeauthoryear{{Drinkwater}, {Webster}, {Francis}, {Condon},
  {Ellison}, {Jauncey}, {Lovell}, {Peterson} \& {Savage}}{{Drinkwater}
  et~al.}{1997}]{dwf+97}
{Drinkwater} M.~J.,  {Webster} R.~L.,  {Francis} P.~J.,  {Condon} J.~J.,
  {Ellison} S.~L.,  {Jauncey} D.~L.,  {Lovell} J.,  {Peterson} B.~A.,
  {Savage} A.,  1997, MNRAS, 284, 85

\bibitem[\protect\citeauthoryear{{Fanti}, {Fanti}, {Schilizzi}, {Spencer}, {Nan
  Rendong}, {Parma}, {van Breugel} \& {Venturi}}{{Fanti} et~al.}{1990}]{ffs+90}
{Fanti} R.,  {Fanti} C.,  {Schilizzi} R.~T.,  {Spencer} R.~E.,  {Nan Rendong}
  {Parma} P.,  {van Breugel} W.~J.~M.,    {Venturi} T.,  1990, A\&A, 231, 333

\bibitem[\protect\citeauthoryear{{Grasha} \& {Darling}}{{Grasha} \&
  {Darling}}{2011}]{gd11}
{Grasha} K.,  {Darling} J.,  2011, in American Astronomical Society Meeting
  Abstracts Vol.~43, {A Search for Intrinsic HI 21-cm Absorption Toward Compact
  Radio Sources}.
p. 345.02

\bibitem[\protect\citeauthoryear{{Isobe}, {Feigelson} \& {Nelson}}{{Isobe}
  et~al.}{1986}]{ifn86}
{Isobe} T.,  {Feigelson} E.,    {Nelson} P.,  1986, ApJ, 306, 490

\bibitem[\protect\citeauthoryear{{Kalberla}, {Dedes}, {Kerp} \&
  {Haud}}{{Kalberla} et~al.}{2007}]{kdkh07}
{Kalberla} P.~M.~W.,  {Dedes} L.,  {Kerp} J.,    {Haud} U.,  2007, A\&A, 469,
  511

\bibitem[\protect\citeauthoryear{{Kalberla} \& {Kerp}}{{Kalberla} \&
  {Kerp}}{2009}]{kk09}
{Kalberla} P.~M.~W.,  {Kerp} J.,  2009, Ann. Rev. Astr. Ap., 47, 27

\bibitem[\protect\citeauthoryear{{Lanfranchi} \& {Fria{\c c}a}}{{Lanfranchi} \&
  {Fria{\c c}a}}{2003}]{lf03}
{Lanfranchi} G.~A.,  {Fria{\c c}a} A.~C.~S.,  2003, MNRAS, 343, 481

\bibitem[\protect\citeauthoryear{{Lavalley}, {Isobe} \& {Feigelson}}{{Lavalley}
  et~al.}{1992}]{lif92a}
{Lavalley} M.~P.,  {Isobe} T.,    {Feigelson} E.~D.,  1992, in BAAS Vol.~24,
  {ASURV, Pennsylvania State University. Report for the period Jan 1990 - Feb
  1992.}.
pp 839--840

\bibitem[\protect\citeauthoryear{{Morganti}}{{Morganti}}{2004}]{mor04}
{Morganti} R.,  2004, in {Bachiller} R.,  {Colomer} F.,  {Desmurs} J.-F.,   {de
  Vicente} P.,  eds, European VLBI Network on New Developments in VLBI Science
  and Technology {Probing the nature of the ISM in Active Galactic Nuclei
  through HI absorption}.
pp 133--141

\bibitem[\protect\citeauthoryear{Murphy, Webb, Flambaum, Drinkwater, Combes \&
  Wiklind}{Murphy et~al.}{2001}]{mwf+00}
Murphy M.~T.,  Webb J.~K.,  Flambaum V.~V.,  Drinkwater M.~J.,  Combes F.,
  Wiklind T.,  2001, MNRAS, 327, 1244

\bibitem[\protect\citeauthoryear{{Obreschkow} \& {Rawlings}}{{Obreschkow} \&
  {Rawlings}}{2009}]{or09a}
{Obreschkow} D.,  {Rawlings} S.,  2009, ApJ, 696, L129

\bibitem[\protect\citeauthoryear{Osterbrock}{Osterbrock}{1989}]{ost89}
Osterbrock D.~E.,  1989, Astrophysics of Gaseous Nebulae and Active Galactic
  Nuclei.
University Science Books, Mill Valley, California

\bibitem[\protect\citeauthoryear{Press, Flannery, Teukolsky \&
  Vetterling}{Press et~al.}{1989}]{pftv89}
Press W.~H.,  Flannery B.~P.,  Teukolsky S.~A.,    Vetterling W.~T.,  1989,
  Numerical Recipes: {T}he Art of Scientific Computing.
Cambridge University Press, Cambridge

\bibitem[\protect\citeauthoryear{Ries}{Ries}{2012}]{rie12}
Ries P.~A.,  2012, PhD thesis, University of Virginia

\bibitem[\protect\citeauthoryear{{Schlegel}, {Finkbeiner} \&
  {Davis}}{{Schlegel} et~al.}{1998}]{sfd98}
{Schlegel} D.~J.,  {Finkbeiner} D.~P.,    {Davis} M.,  1998, ApJ, 500, 525

\bibitem[\protect\citeauthoryear{{Stanghellini}, {Dallacasa}, {O'Dea}, {Baum},
  {Fanti} \& {Fanti}}{{Stanghellini} et~al.}{2001}]{sdo+01}
{Stanghellini} C.,  {Dallacasa} D.,  {O'Dea} C.~P.,  {Baum} S.~A.,  {Fanti} R.,
     {Fanti} C.,  2001, A\&A, 377, 377

\bibitem[\protect\citeauthoryear{Vermeulen, Pihlstr\"{o}m, Tschager, de Vries,
  Conway, Barthel, Baum, Braun, Bremer, Miley, O'Dea, Roettgering, Schilizzi,
  Snellen \& Taylor}{Vermeulen et~al.}{2003}]{vpt+03}
Vermeulen R.~C.,  et al.,  2003, A\&A, 404, 861

\bibitem[\protect\citeauthoryear{{Waldram}, {Yates}, {Riley} \&
  {Warner}}{{Waldram} et~al.}{1996}]{wyrw96}
{Waldram} E.~M.,  {Yates} J.~A.,  {Riley} J.~M.,    {Warner} P.~J.,  1996,
  MNRAS, 282, 779

\bibitem[\protect\citeauthoryear{{Wall} \& {Peacock}}{{Wall} \&
  {Peacock}}{1985}]{wp85}
{Wall} J.~V.,  {Peacock} J.~A.,  1985, MNRAS, 216, 173

\end{thebibliography}

\section*{APPENDIX --- SED fitting }
\label{appa}

In order to estimate the radio spectral indices and turnover frequencies, as well as the $\lambda\leq912$ \AA\ continuum
luminosities and photon rates, we produced an SED for each source by compiling the NED photometries, the Galaxy
Evolution Explorer (GALEX) near and far ultra-violet fluxes, as well as the $BVRK$ magnitudes, published elsewhere
(compiled in \citealt{cww+08,cw10}). 
After correcting for Galactic extinction using the maps of \citet{sfd98},
for each of the 241 sources the flux density was converted to a luminosity and the frequency de-redshifted to the source rest-frame.
\begin{figure}
\vspace{6.0cm} 
	\setlength{\unitlength}{1in} 
\begin{picture}(0,0)
\put(-0.2,2.85){\includegraphics{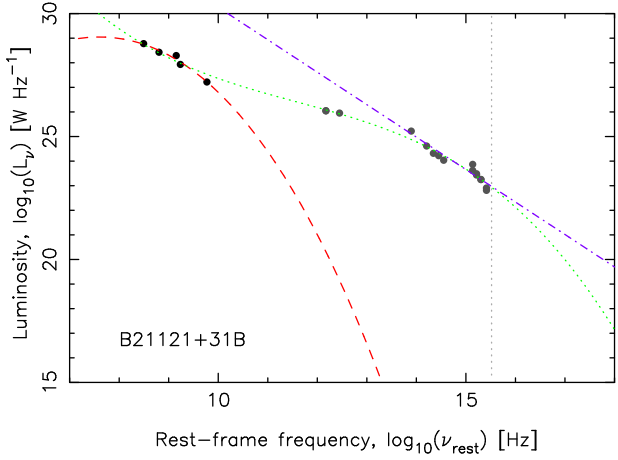}}
\end{picture}
\caption{SED fits to B2\,1121+31B. 
As coded in Fig. \ref{SEDs}, the curves show the polynomial fits to the radio ($\nu_{\rm rest}
  \leq10^{12}$ Hz) data (dashed)  and  to the full ($\nu_{\rm rest} \leq 10^{17}$ Hz) data (dotted). The straight dot-dashed line
shows the power-law fit to the $10^{14.5} \leq \nu_{\rm rest} \leq 10^{17}$ Hz 
data. The vertical dotted line highlights $\nu_{\rm rest} = 3.29 \times10^{15}$ Hz ($\lambda = 912$ \AA).}
\label{J0003+2129}
\end{figure}
\begin{figure}
\vspace{6.0cm} 
	\setlength{\unitlength}{1in} 
\begin{picture}(0,0)
\put(-0.2,2.85){\includegraphics{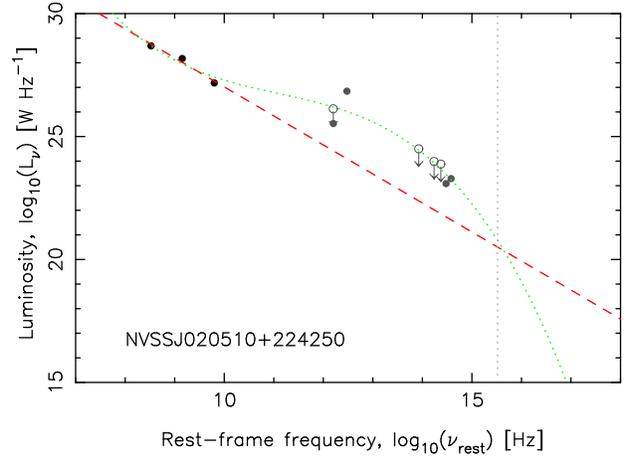}}
\end{picture}
\caption{SED fits to NVSS\,J020510+224250 demonstrating a case where no radio turnover is apparent and where the
  $\lambda=912$ \AA\ continuum luminosity cannot be determined (cf. Fig.~\ref{J0003+2129}).}
\label{3C153}
\end{figure}
 
Using Vandermonde matrices \citep{pftv89}, three polynomials were fit to the data points for each source:
\begin{enumerate}
\item {\em The radio SED}: A  third or second order polynomial was fit to the $\nu_{\rm rest}\leq10^{12}$ Hz points.
 The turnover frequency was then obtained from $dL_{\nu}/d\nu =0$ (with a  maximum being tested using
  $d^2L_{\nu}/d\nu^2 <0$) and the spectral index from the value of $\alpha= dL_{\nu}/d\nu$ at $\nu_{\rm rest}
  = 1.42\times10^9$ Hz, thus being defined as $S\propto\nu^{\alpha}$.
     Fig. \ref{J0003+2129} shows an example of a possible turnover (at $40$ MHz) and a spectral index of  $\alpha= -1.21$.
     In the case where no radio turnover is apparent, providing $dL_{\nu}/d\nu$ is negative in the
     radio-band, the lowest observed frequency is assigned as an upper limit to this (as in Fig. \ref{3C153}). That is,
     we assume that the turnover occurs at a frequency lower than that currently observed \citep{ffs+90}.  Otherwise, if
     $dL_{\nu}/d\nu\geq0$, the source is flagged as having an indeterminable turnover frequency.

\item {\em The full  SED}:  A third order polynomial was fit to the all of the data points at $\nu_{\rm rest}\leq10^{17}$ Hz. This
fit provides a visual guide only.

 \item {\em The  UV SED}: A first order polynomial (power-law) was fit to the $10^{14.5} \leq \nu_{\rm rest} \leq 10^{17}$ Hz data.
   This upper cut-off is applied since, although relevant and used to derive the composite SEDs \citep{cw12}, the, usually sparsely sampled,
 X-ray points can have a strong influence on the fit, although these may
   be arising from other mechanisms than those producing the UV SED:
   \begin{enumerate} 
  \item Provided that $dL_{\nu}/d\nu<0$ (e.g. as in Fig. \ref{J0003+2129}), 
       the UV SED was used to derive the luminosities and photon rate.

   
     \item Should the gradient exceed $dL_{\nu}/d\nu=0$ at $\nu_{\rm rest} =
       3.29\times10^{15}$ Hz, usually from a concentration of data points,  the luminosity is estimated from the mean of any neighbouring data points (at $ \nu_{\rm
         rest} = 10^{15.52\pm0.50}$ Hz). 

\end{enumerate}
Only in the first case (a), where a reliable fit is obtained over the range specified, is the ionising photon rate derived from
\[
\int^{\infty}_{\nu}\frac{L_{\nu}}{h\nu}\,d{\nu},~{\rm where}~\log_{10}L_{\nu} = \alpha\log_{10}\nu+ {\cal C} \Rightarrow  L_{\nu} = 10^{\cal C}\nu^{\alpha}
\]
for a power-law fit, where \AL\ is the spectral index and ${\cal C}$ the intercept. This gives,
\[
\frac{10^{\cal C}}{h}\int^{\infty}_{\nu}\nu^{\alpha-1}\,d{\nu} = \frac{10^{\cal C}}{\alpha h}\left[\nu^{\alpha}\right]^{\infty}_{\nu} = \frac{-10^{\cal C}}{\alpha h}\nu^{\alpha}~{\rm where}~\alpha < 0,
\]
otherwise the ionising photon rate is flagged as indeterminable.
\end{enumerate}

\begin{figure*}
\centering \includegraphics[angle=-90,scale=0.75]{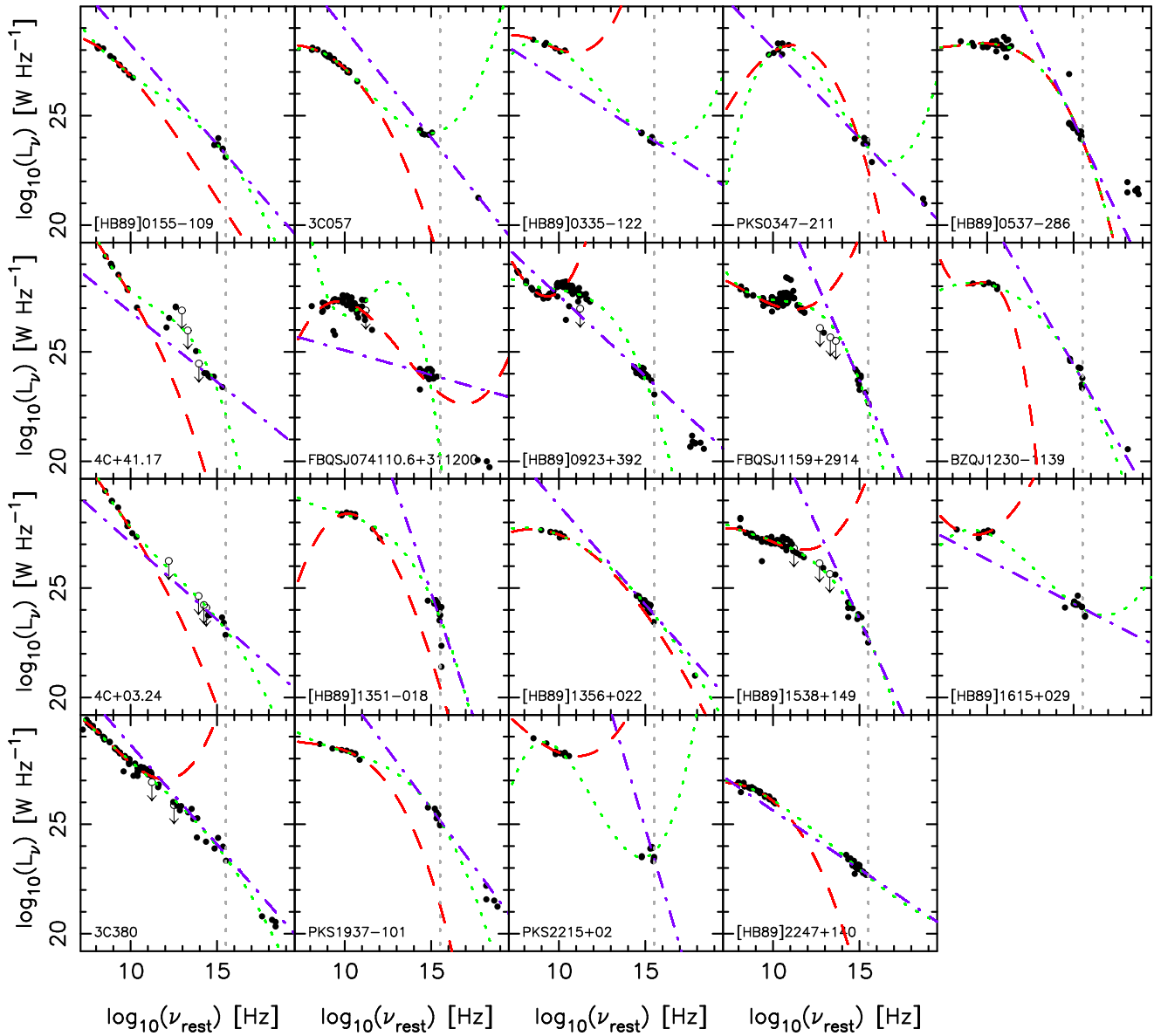}
\caption{As per Fig. \ref{SEDs}, but for the 19 UV luminous sources \citep{cwm+10}.}
\label{SEDs_2}
\end{figure*}

Finally, in Fig. \ref{SEDs_2} we show the SED fits to the UV luminous sources (those which lie above the
$L_{\rm UV}=10^{23}$ \WpHz\ cut-off, the last row in Table \ref{stats}).
As seen from this, the ultra-violet luminosities may be estimated and photon rates derived for all of these (see \citealt{cw12}).

\label{lastpage}

\end{document}